\documentclass[journal]{IEEEtran}
\ifCLASSINFOpdf
\else
\fi

\usepackage{stackengine}

\usepackage{graphicx}
\usepackage[resetlabels,labeled]{multibib}
\usepackage{multibib}
\usepackage{array}

\newcites{A}{Research has an Empirical Component which Studies Trade-offs or like Decisions}
\newcites{B}{Research has an Empirical Component \& Discusses Trade-offs or like Decisions}
\newcites{C}{Research has an Empirical Component \& Does Not Discuss Trade-offs or like Decisions}
\newcites{D}{Research Does Not Have an Empirical Component \& Discusses Trade-offs or like Decisions}
\newcites{E}{Research Does Not Have an Empirical Component \& Does not Discuss Trade-offs or like Decisions}

\begin{document}
%
\title{A Systematic Literature Review on \\ Intertemporal Choice in Software Engineering -- \\ Protocol and Results}
%
%
%

\author{Christoph~Becker,
        Dawn~Walker,
        and Curtis McCord
\thanks{C. Becker is lead of the Digital Curation Institute and with the Faculty of Information, University of Toronto, Toronto, ON, Canada. E-mail: (see https://ischool.utoronto.ca/faculty/christoph-becker).}
\thanks{D. Walker and C. McCord are doctoral students at the Faculty of Information, University of Toronto, Toronto, ON, Canada}
\thanks{Submission on January 27, 2017.}}

%
%

\maketitle

\begin{abstract}
When making choices in software projects, engineers and other stakeholders engage in decision making that involves uncertain future outcomes. Research in psychology, behavioral economics and neuroscience has questioned many of the classical assumptions of how such decisions are made.

This literature review aims to characterize the assumptions that underpin the study of these decisions in Software Engineering. We identify empirical research on this subject and analyze how the role of time has been characterized in the study of decision making in SE.

The literature review aims to support the development of descriptive frameworks for empirical studies of intertemporal decision making in practice.
\end{abstract}


\IEEEpeerreviewmaketitle

\section{Introduction}
\IEEEPARstart{C}{omplex} software-intensive systems play critical roles in our societies: their ongoing development, innovation, and maintenance is intertwined with our everyday social and economic activities. As recognition of the key role software technology can play in society's sustainability grows, the need for a paradigm shift in the mindset of the software industry has become clear. Sustainability is often defined within the domain of ``sustainable development'', which ``meets the needs of the present generation without compromising the ability of future generations to meet their own needs'' \cite{wced1987ocf}. At its core, sustainability is the capacity to endure, but sustainability of social systems is different than technical or natural systems. Originally equated with environmental concerns, it is now clear that sustainability requires equal consideration of five dimensions: environmental, societal, individual, economic, and technical \cite{penzenstadler2014safety}. 

Within and across these concerns, software engineering (SE) decisions are made about system scope, goals and objectives, features, functions, architectural designs, and many other areas throughout the development lifecycle. The effects of these choices are often delayed, and many critical decisions involve trade-offs between outcomes at different points in time. In such cases, longer-term consequences are not always sufficiently considered \cite{neumann2012foresight,becker2016requirements}. 

Research in psychology and behavioral economics calls choices that involve trade-offs across time ``intertemporal choices,'' defining them as ``decisions involving tradeoffs among costs and benefits occurring at different times'' \cite{frederick2002time}. Researchers have developed a number of theories of such choices \cite{frederick2002time,loewenstein2003time} and have demonstrated that straightforward assumptions about how decision makers evaluate and discount the future are often misguided and wrong \cite{frederick2002time,loewenstein2003time}.
 
Herein,an important distinction is made between normative and descriptive decision theories. Normative theories focus on the identification of the best decision, and model an ideal decision maker. Normative models of how decisions are made in SE commonly assume a rational agent (with reasonable cognitive boundaries) choosing between a set of options according to a value function. 

From choosing a software development methodology to evaluating release planning, prioritizing requirements and choosing between architectural design options, SE literature commonly assumes that decision making operates in a predictable, rational way. For example, one author writes ``In most problems, to make a decision, a situation is assessed against a set of characteristics or attributes, also called criteria. Decision making based on various criteria is supported by multi-criteria methodologies'' \citeB{simao2016task}. The assumption is that a team of competent engineers evaluates the options to the best of their knowledge, and they choose the option with the highest expected value. Much of their discussion in theory and practice focuses on how to best estimate that value. The frameworks of Value Based Software Engineering aim to base SE decisions more explicitly on an understanding of value \cite{biffl2006value}. Most commonly, this value is expressed in economic terms, and the incommensurability of multiple aspects of value is often addressed through application of utility functions \cite{keeney1993decisions}.

The theory of expected utility stems from game theory \cite{von1944theory} and was developed from principles, not empirical study. By contrast, descriptive theories aim to characterize the behavior of actual decision making. As Tversky and Kahneman write, ``The modern theory of decision making under risk emerged from a logical analysis of games of chance rather than from a psychological analysis of risk and value. The theory was conceived as a normative model of an idealized decision maker, not as a description of the behavior of real people... the logic of choice does not provide an adequate foundation for a descriptive theory of decision making'' \cite{tversky1986rational}.

Criticism of the prevailing normative decision theories has come from numerous perspectives, and various alternative conceptions have been proposed. For example, well-known experiments have shown that people do not discount the future linearly \cite{frederick2002time} and that risk aversion is higher for gains than for losses \cite{tversky1986rational}. More substantively, Tversky and Kahneman showed that some of the foundational axioms of normative decision theory, and in particular expected utility theory, are inconsistent with observable behavior. More radically, Klein's study of expert decision making showed convincingly that experienced decision makers do not actually weigh a set of alternatives against criteria to maximize expected utility when making critical choices \cite{klein1999sources}. It is this divergence between prevalent normative models and observed behavior that motivated this review. This corresponds to the recent emergence of Behavioral Software Engineering, a field that aims to draw in behavioural frameworks and concepts for a better understanding of software engineering \cite{lenberg2014towards,lenberg2015behavioral,hofman2011behavioral}.

In SE, choices that are expedient in the short-term but create unwanted longer-term consequences have been conceptualized most prominently as `technical debt', which focuses on engineering choices that create hidden costs. The metaphor of debt aims to make these hidden costs visible and manageable. Interpreted more broadly, the notion of `sustainability debt' expands the metaphor to direct and indirect effects across all dimensions of sustainability \cite{betz2015sustainability}.

\subsection{Objective}

This review is motivated by the need to better understand how and why software practitioners incur sustainability debt in practice. In order to develop a descriptive framework for intertemporal choices in SE, we review the literature to identify whether the intersection of these concepts has been acknowledged and addressed; describe which perspectives and assumptions about decision makers underpin existing research; and analyze how the role of time has been characterized in the study of decision making in SE.

Because of our interest in distinguishing normative models theorizing about decision-making in SE from descriptive, empirical accounts of how trade-off decisions relating to time are made in software design projects, we will first map empirical and other types of research of decision making, and then analyze empirical research in detail in order to understand the assumptions of decision making models that underpin this research.

\subsection{Contribution}

We aim to reveal how trade-off choices have been conceptualized within SE so far, identify gaps in how decision making is reviewed and investigated, and map how SE literature approaches making trade-offs over time. 

\section{Literature Review Study Design}

\subsection{Research Questions}

We characterize perspectives on decision making within SE research through the following questions:
\begin{enumerate}
	\item[RQ1] Which empirical research in SE has studied trade-off decisions involving time?
	\item[RQ2] Which dimensions are considered in these studies?
	\item[RQ3] How has the role of time been conceptualized in these studies?
	\item[RQ4] Which assumptions on decision making underpin these studies?
\end{enumerate}

While we are interested in the assumptions on decision making that underpin the perspective of the non-empirical studies, we focus our in-depth analysis on empirical work due to time restrictions.

\subsection{Roles and Responsibilities}

The roles and responsibilities for this project are defined in Table \ref{roles}. We have one principal researcher, Christoph Becker, and two supporting researchers, Curtis McCord and Dawn Walker. External reviews were conducted by Stefanie Betz and Ruzanna Chitchyan.

\begin{table*}[t]
\centering
\caption{Breakdown of Roles and Responsibilities}
\label{roles}
\begin{tabular}{|l|l|c|c|l|}
\hline
                                               & Christoph Becker       & \multicolumn{1}{l|}{Curtis McCord} & \multicolumn{1}{l|}{Dawn Walker} & External Reviewers (Betz, Chitchyan) \\ \hline
Develop Protocol                               & \multicolumn{1}{c|}{X} & X                                  & X                                &                                      \\ \hline
Prototype Protocol                             &                        & \multicolumn{1}{l|}{}              & X                                &                                      \\ \hline
Define Search Strings                          & \multicolumn{1}{c|}{X} & X                                  & \multicolumn{1}{l|}{}            &                                      \\ \hline
Define Classification Scheme                   & \multicolumn{1}{c|}{X} & \multicolumn{1}{l|}{}              & X                                &                                      \\ \hline
Review of Protocol                             & \multicolumn{1}{c|}{X} & \multicolumn{1}{l|}{}              & \multicolumn{1}{l|}{}            & \multicolumn{1}{c|}{X}               \\ \hline
Final Revision of Protocol                     &                        & X                                  & X                                &                                      \\ \hline
Identify Primary Research                      &                        & X                                  & X                                &                                      \\ \hline
Retrieve Primary Research                      &                        & X                                  & X                                & `                                    \\ \hline
De-duplicate                                        &                        & X                                  & \multicolumn{1}{l|}{}            &                                      \\ \hline
Prototype Relevancy Voting                     & \multicolumn{1}{c|}{X} & X                                  & X                                &                                      \\ \hline
Relevancy Voting                               &                        & X                                  & X                                &                                      \\ \hline
Review of Relevancy Vote                       & \multicolumn{1}{c|}{X} & \multicolumn{1}{l|}{}              & \multicolumn{1}{l|}{}            &                                      \\ \hline
Data extraction, Classification, and Synthesis &                        & X                                  & X                                &                                      \\ \hline
Analysis Validation                            & \multicolumn{1}{c|}{X} & \multicolumn{1}{l|}{}              & \multicolumn{1}{l|}{}            &                                      \\ \hline
Write Technical Report                         &             \multicolumn{1}{c|}{X}        & X                                  & X                                &                                      \\ \hline
Review of Technical Report                     & \multicolumn{1}{c|}{X} & \multicolumn{1}{l|}{}              & \multicolumn{1}{l|}{}            & \multicolumn{1}{c|}{X}               \\ \hline
\end{tabular}
\end{table*}

\subsection{Search Strategy}

In order to produce a systematic overview of this area, the overall search process for this literature review is based on guidelines established by Kitchenham \cite{kitchenham2004procedures}.

\subsubsection{Information Sources}

We performed automated searches on the following indexing systems and digital libraries: Scopus, IEEE Xplore, and ACM Digital Library.

\subsubsection{Preliminary search}

The term `intertemporal choice' has come to describe precisely our area of interest. At an early stage, we conducted searches to identify whether there has been explicit attention to this concept in the literature. 

\textbf{``intertemporal choice'' \\
\indent AND ``software engineering''}

\begin{figure}
\caption{Overlap between SE and Intertemporal Choice SCOPUS query}
\label{ic-mq}
  \centering
    \includegraphics[scale=0.15]{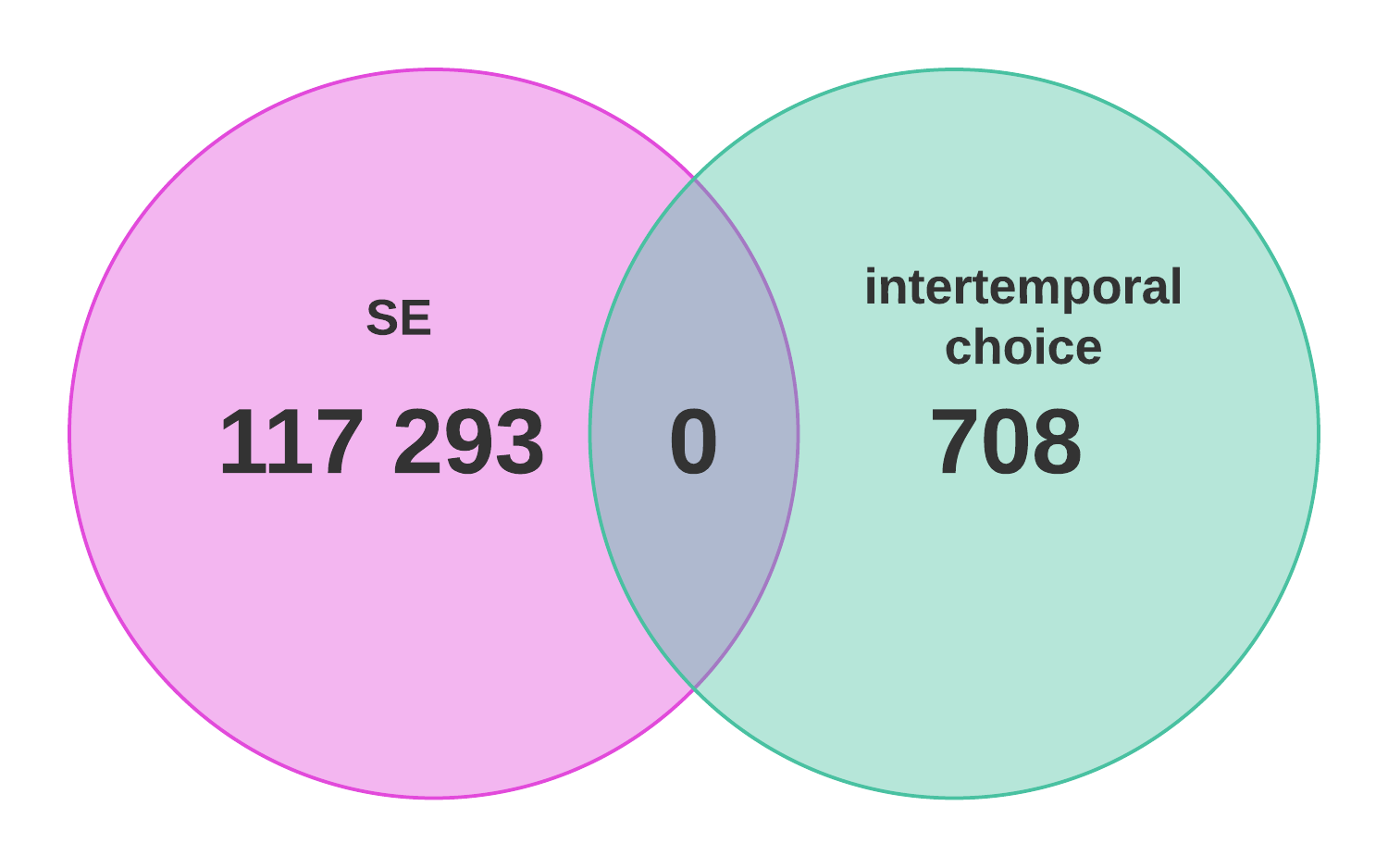}
\end{figure}

\begin{table}
\centering
\caption{Overview of SCOPUS search results for Intertemporal Choice preliminary search}
\label{ic-overview}
\begin{tabular}{|l|l|}
\hline
Total number of search results & 0 \\ \hline
\end{tabular}
\end{table}

Fig. \ref{ic-mq} and Table \ref{ic-overview} show the resulting search numbers for SCOPUS. While the exact numbers differ for the other databases, the trend is mirrored and the intersection remained empty for all searches.

The search revealed that intertemporal choice is not explicitly treated in the literature at all, and the phrasing was not present in any papers. This does not necessarily indicate that SE does not deal with intertemporal choices, but the absence of explicit mention of the term ``intertemporal choice'' suggests that the concepts arising from the field of behavioral economics have not been congruently linked to SE, i.e. that no direct conceptual mapping has been established between the two disciplines yet.

\subsubsection{Preliminary concept review}

Before conducting further searches, we aimed to establish a candidate set of concepts that would scaffold our understanding of decision making vocabulary. To do so, we reviewed textbooks and standards in Software Engineering \cite{ieee2013swebok}, Value Based Software Engineering \cite{biffl2006value}, Decision Analysis, Behavioural Economics \cite{cartwright2014behavioral} and Management Theory \cite{cooper1997blackwell} to compose a working vocabulary of terms related to intertemporal choice. From these texts we developed a series of prototypical concept maps that decomposed key components of decision making into potential search terms. Terms such as ``cost'', ``value'', ``benefit'', ``risk'', ``decision-making'', for example, were widely used across disciplines, and helped to structure our understanding of decisions and provide terminology for coding and analysis later on.

\subsubsection{Search String}

The goal for the search string was to capture results that dealt with intertemporal decision-making in SE, to examine how SE projects saw time as a factor in their decision processes, how they make decisions about the future of their projects, and how they might weigh future and present goods against each other. We included the clause ``software engineering'' to limit the disciplinary scope of our research-- other disciplinary scopes such as ``requirements engineering'' could conceivably lead to different perspectives.

To capture the temporal aspect of decision making, we settled on the general term of ``time'', with the intention of using more specific coding during analysis. Preliminary search queries (See Appendix \ref{long-strings}) were more complex and used more discipline-specific jargon (``life cycle'', ``endurance'') whose specificity would occlude relevant papers that could be captured by a more general query.

While these searches included relevant results that connected to the concepts that emerged from initial review, the results were mixed and widely spread across disciplines. It became clear that introducing divergent and specific terms from multiple disciplines would increase the amount of papers captured, but not necessarily make the literature review more effective or representative. This, and the possible bias introduced through these more complex queries, led us to choose a simplified more generic query string and move some of the detailed aspects of intertemporal choice to the coding and analysis stages.

The same reasoning process governed our decisions on the second clause of our query; we were interested in papers discussing trade-offs, but recognized that while the term is widely used, it might not be used by all authors describing these types of decisions. What we really wanted to capture through coding was choices that required parties to weigh decision dimensions against each other.
 
Search queries were piloted twice (See Appendix \ref{pretest-strings} for pretest queries) prior to establishing the final search string: 
\indent \textbf{time \\
\indent AND ``decision making'' \\
\indent AND ``software engineering''}

\subsubsection{Ancillary Search}

Using the same search strategy, one ancillary search was performed as part of the literature review:
\indent\textbf{``technical debt''}

The concept of Technical Debt (TD) is prominent in software engineering and closely related to the dimensions of our main query. Technical debt can be defined as: ``a design or construction approach that is expedient in the short term but that creates a technical context in which the same work will cost more to do later than it would cost to do now (including increased cost over time)'' (Ernst \cite{ernst2015measure}, borrowing from McConnell). In this framing, TD always includes an explicitly temporal dimension, built into the concept of debt. Decisions that are made about TD would presumably include a temporal dimension and the commensuration of future and present goods. As such, the literature on TD could be complementary to other areas of intertemporal choice and shed light on specific assumptions.

The results of the ancillary query were documented, but the only analysis performed within this review was an identification of the overlap with the primary search, as described further below (see Section \ref{results}). The resulting corpus of publications will be used for further analysis in the future.

\subsection{Selection Criteria}

\subsubsection{Inclusion Criteria}

We established the following criteria to identify relevant publications that would answer research questions:

\begin{itemize}
\item \textbf{Publication Year}: All years were included.
\item \textbf{Publication Type}: We included peer-reviewed papers published in journals, conference proceedings, and workshop proceedings.
\item \textbf{Content}: The paper had to contain a discussion of decision-making in software engineering projects.
\item \textbf{Coverage}: The paper had to cover development of a software system rather than only hardware.
\end{itemize}

\subsubsection{Exclusion Criteria}

\begin{itemize}
\item \textbf{Publication Language}: We excluded papers in languages other than English.
\item \textbf{Publication Quality}: We excluded papers retracted by the publisher.
\item \textbf{Publication Type}: We excluded non-paper results including: posters, abstract-only submissions, book reviews, books, entire volumes of proceedings, panels, presentations, tutorials, opinion pieces.
\item \textbf{Technical}: We excluded papers where the PDF was unavailable (behind a paywall or not locatable).
\end{itemize}

\subsection{Selection Procedures}

After downloading, removing duplicates, and applying our exclusion criteria, the remaining papers were screened for relevancy using the following procedure:

\begin{enumerate}
\item The secondary researchers voted on relevance: They read identified paper titles and abstracts in order to decide on inclusion using the criteria above. A yes or no decision (``Y/N'') was assigned as well as a certainty value from 1-3 (where 3=certain).
\item Voters reviewed 10 of the 307 papers as a pilot, and then conducted a larger pilot of 49 papers including the original 10, and discussed the results together.
\item Following this quality assurance step, the remaining 258 papers were split and reviewed by one voter each, following the voting process established above.
\item In cases where papers were reviewed by more than one voter, disagreements were resolved through discussion and consensus.
\item All decisions on papers reviewed by only one voter were compiled. Those with a certainty value below 2 (191) were reviewed and discussed by both voters, and a randomly selected sample of 65 was evaluated for consistency. In case of remaining doubt, the papers were included.
\item 155 papers marked for further coding were looked over by the an internal reviewer to determine whether inclusion and exclusion criteria were appropriate before analysis. Because of the rule to include papers in case of doubts, the focus was on verifying included papers at this stage.
\end{enumerate}

\subsection{Analysis}

Papers from the relevancy review were analyzed in order to address the research questions established above. Application of a checklist as well as review ensured the quality of analysis and coding. Researchers extracted data using a form to capture fields relevant to our research questions.

\subsubsection{Assessment}

In order to ensure quality of analysis and findings, researchers conducted multiple internal reviews throughout many stages of the Literature Review: protocol, relevancy voting procedure, relevancy voting results, technical report. Additional external review of the technical report led to suggestions for improvement.

\subsubsection{Data Extraction}
Researchers classified studies according to the type or domain of the decision-making studied, the methods of investigation and research, whether there was a trade-off decision, and if so, the dimensions of the trade-off. Free annotation was also used to capture additional information the coders deemed relevant.

A form was also designed to capture these fields as well as metadata from the studies, including author, title, year of publication, DOI, and unique document key (generated from author, title, publication).\\
\textbf{Coding Field 1: Scope of the Decision}
\begin{description}
\item [(pm)] Project Management \cite{ieee2013swebok}
\item [(dev)] Software Requirements, Design, Architecture, Development \cite{ieee2013swebok}
\item [(mait)] Software Maintenance
\item[(other)] Including business and business strategy
\end{description} 

Within SE, requirements decisions, design, architecture, and development cover a wide range of tasks, fields, etc. We intentionally grouped these together to cover all engineering decisions as one, in part because these decisions often span multiple areas.

\textbf{Coding Field 2: Research Methodology}
\begin{description}
\item[(emp)] Empirical methods were used and the object of empirical study was a decision
\item[(emp comp)] \hfil \\ {Empirical methods were used and the object of empirical study was NOT a decision}
\item[(lit)] The paper was (exclusively) a literature review or a systematic mapping study
\item [(other)] The research was not empirical, i.e. it was theoretical or attempted to develop a model 
\end{description}
\textbf{Coding Field 3: Does the article discuss trade-off decisions?}
\begin{description}
\item[(Y)] Yes
\item[(N)] No
\end{description}
\textbf{Coding Field 4: Dimensions of Trade-Off}
\begin{description}
\item[(cost)] Cost (Often in monetary terms)
\item[(func)] Functionality
\item[(mait)] Software Maintainability
\item[(qual)] Software Quality
\item[(risk)] Risk
\item[(time)] Time (Includes scheduling, delivery and release)
\item[(value)] Value (As in terms of monetary value, ``business value'' or in some cases, in terms of benefit)
\item[(other)] See Appendix 
\end{description}

\begin{figure}
\caption{Overlap between main query and TD papers}
\label{td-mq}
  \centering
    \includegraphics[scale=0.15]{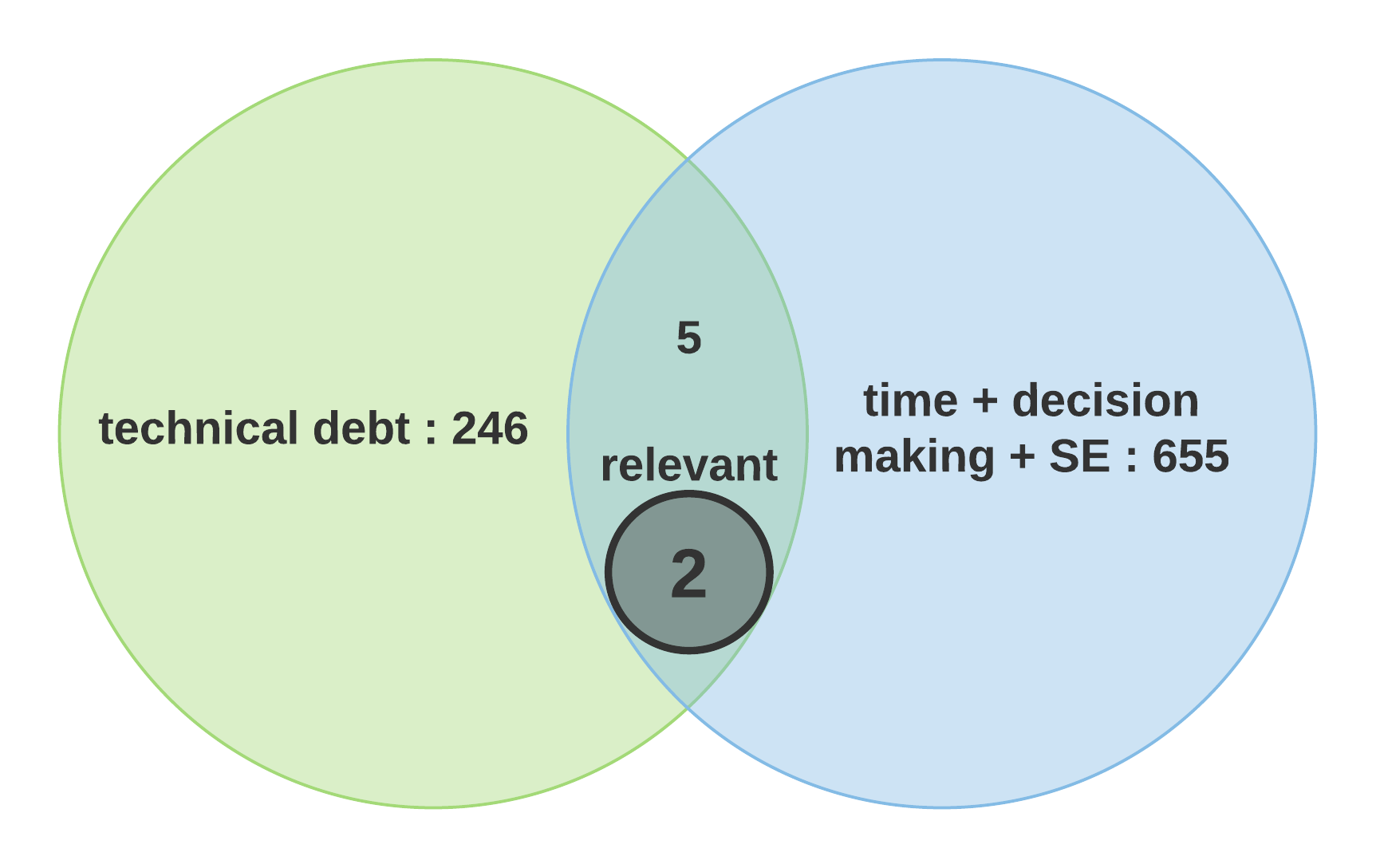}
\end{figure}

\begin{table}
\centering
\caption{Overview of search results for main search}
\label{main-search-overview}
\begin{tabular}{|p{2.5in}|l|}
\hline
Total Number of Search Results                                                 & 889 \\ \hline
Total number of results after duplicate removal and exclusion criteria applied & 652 \\ \hline
Number selected after preliminary relevancy review                             & 307 \\ \hline
Number selected after voting                                                   & 155 \\ \hline
\end{tabular}
\end{table}

\begin{table}
\centering
\caption{Overview of search results for Technical Debt ancillary search}
\label{td-overview}
\begin{tabular}{|p{2.5in}|l|}
\hline
Total number of search results                                                 & 620 \\ \hline
Total number of results after duplicate removal and exclusion criteria applied & 246 \\ \hline
\end{tabular}
\end{table}

\subsubsection{Analysis of Extracted Data}

The secondary researchers extracted data and analyzed the results included below. From this they synthesized findings on the current research. Feedback was provided through an internal review by the principal researcher. In order to analyze the extracted data, the researchers:

\begin{itemize}
\item derived statistics of coded categories for mapping extracted data
\item mapped out areas of existing work
\item created visualizations with groups of dimensions
\end{itemize}

\section{Results}
\label{results}

Search result statistics are provided in Table \ref{main-search-overview}. First the search results from the indexing systems and digital libraries were compiled, then results were de-duplicated and exclusion criteria applied. From those 652 papers, and initial assessment to determine whether they were relevant led to 307 papers selected. Based on voting and discussion to reach consensus, that number was reduced to 155 for final coding.

Statistics of the ancillary ``Technical Debt'' query are summarized in Table \ref{td-overview}. As expected, there was some overlap between the papers returned in our Technical Debt query and those returned in our main query. As technical debt has become a more prominent term in software engineering discourse, it also becomes a phenomenon which can be analyzed and accounted for. In this way it becomes manageable: the object of decision-making. For these reasons we were not surprised to find several recent papers that focus on decisions about monitoring, reporting and managing technical debt. Before relevancy voting, 7 papers were in both the technical debt and main query corpus. After relevancy voting, the intersection of the two corpuses was 2 papers, as illustrated in Fig. \ref{td-mq}: 
\begin{itemize}
\item Martini and Bosch, 2016, \textit{An Empirically Developed Method to Aid Decisions on Architectural Technical Debt Refactoring: AnaConDebt} \cite{martini2016empirically}
\item Oliveira, Goldman, and Santos, 2015, \textit{Managing Technical Debt in Software Projects Using Scrum: An Action Research} \cite{oliveira2015managing}
\end{itemize}

\section{Findings}

To explore assumptions that underpin the existing empirical work on trade-off decisions in SE, we first extracted statistics from the coded categories of each paper in order to map extracted data. Subsequently, areas of existing empirical work were further identified, visualized, and data on the groups of dimensions was collated. From this, final analysis was performed in depth on the subset of empirical papers that discussed decision making in SE. We will discuss the main research questions in separate sections below.

\subsection{Which empirical research in SE has studied trade-off decisions involving time?}

As described above, the relevant publications were coded for an empirical focus on studying decisions, and for including in particular decisions involving `trade-offs'. 

\begin{figure}
\caption{Segments distinguished according to research type and focus}
\label{mq-breakdown}
  \centering
    \includegraphics[scale=0.15]{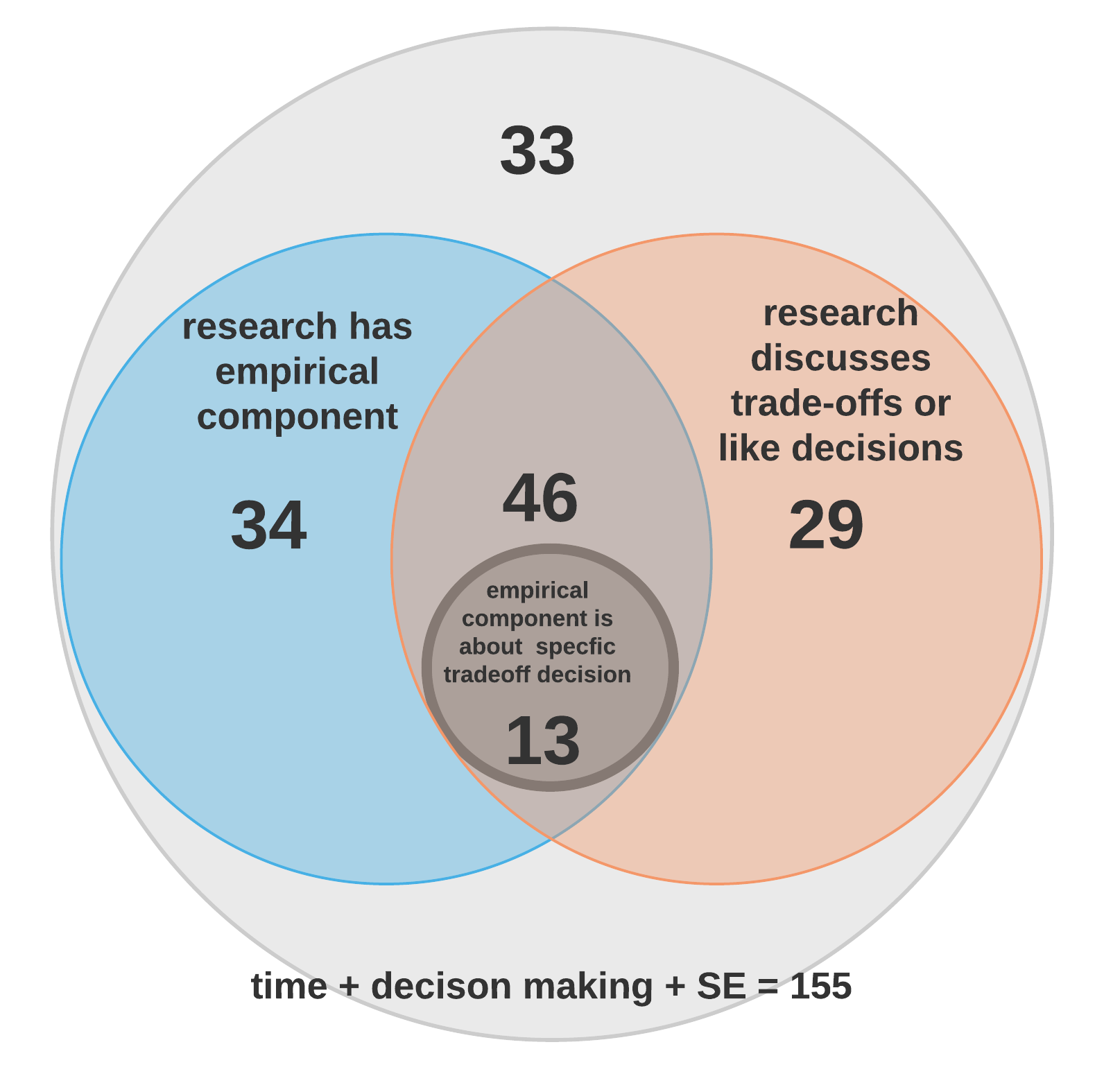}
\end{figure}

\begin{table*}[t]
\centering
\caption{Arrangement of Selected Papers by Type (see Appendix \ref{selected-papers})}
\label{my-label}
\begin{tabular}{|l|l|l|}
\hline
Segment     & Count & Description                                                                                  \\ \hline
{[}A1-13{]} & 13    & Research has an Empirical Component which Studies Trade-Offs or Like Decisions                                         \\ \hline
{[}B1-46{]} & 46    & Research has an Empirical Component and Discusses Trade-Offs or Like Decisions               \\ \hline
{[}C1-34{]} & 34    & Research has an Empirical Component and Does Not Discuss Trade-Offs or Like Decisions        \\ \hline
{[}D1-29{]} & 29    & Research Does Not have an Empirical Component and Discusses Trade-Offs or Like Decisions     \\ \hline
{[}E1-33{]} & 33    & Research Does Not have Empirical Component and Does Not Discuss Trade-Offs or Like Decisions \\ \hline
\end{tabular}
\end{table*}

We found 93/155 papers had some degree of empirical component, and 88/155 discussed trade-offs in such capacities. The resulting Venn diagram shown in Fig. \ref{mq-breakdown} shows a quite even distribution across the emerging subsegments, but indicates that only 13 studies were identified that explicitly used empirical methods to study trade-off decisions where time was a relevant element. This set represents papers that attempt to examine decision-making in software engineering in real or experimental situations. 

Table \ref{emp-chars} summarizes key characteristics of the 13 identified papers, including the research method(s) and citation counts. The most prominent method is case study research.

\begin{table*}[t]
\centering
\caption{Empirical Studies (chronological)}
\label{emp-chars}
\begin{tabular}{|p{3.8cm}|p{0.6cm}|p{2.5cm}|p{7.3cm}|p{1.4cm}|}
\hline
Title & Year & Author(s) & Research Method and Summary & Citation Count (GS) \\ \hline
A cost-value approach for prioritizing requirements \cite{karlsson1997cost} & 1997 & J. Karlsson and K. Ryan & The authors developed a cost–value approach for prioritizing requirements and applied it to two commercial projects (\textbf{case study} evaluation). & 715 \\ \hline
Evaluating the cost of software quality \cite{slaughter1998evaluating} & 1998 & S. A. Slaughter, D. E. Harter, and M. S. Krishnan & The paper analyzes large-scale data about software quality and costs collected empirically from software organizations across time. No individual projects or decisions are studied directly. & 244 \\ \hline
The impact of goals on software project management: An experimental investigation \cite{abdel1999impact} & 1999 & T. K. Abdel-Hamid, K. Sengupta, and C. Swett & An \textbf{experiment} was performed with a project simulation game played in teams. Two control groups played the same game but with different goals-- one focused on minimizing cost and schedule, one delivering highest quality in minimal schedule. The focus was not on these dimension however, but on the role of goal setting in performance. & 148 \\ \hline
Measuring the ROI of software process improvement \cite{van2004measuring} & 2004 & R. Van Solingen & Two \textbf{cases} of (real) projects are described as part of an argument about the need to estimate value so that cost and value can be used to estimate ROI of process improvement. & 120 \\ \hline
A quality-driven systematic approach for architecting distributed software applications \cite{al2005quality} & 2005 & T. Al-Naeem, I. Gorton, M. A. Babar, F. Rabhi, and B. Benatallah & A \textbf{case study} is conducted as part of the evaluation. Interviews with the architect of a real system are conducted to identify decisions. An approach is proposed and its applicability is discussed using the scenario of that real system. The architect was asked for feedback. & 105 \\ \hline
What is important when deciding to include a software requirement in a project or release? \cite{wohlin2005important} & 2005 & C. Wohlin and A. Aurum & A \textbf{questionnaire} was used to identify types of criteria that are most important for requirements prioritization based on industry responses. & 57 \\ \hline
A qualitative empirical evaluation of design decisions \cite{zannier2005qualitative} & 2005 & C. Zannier and F. Maurer & The paper proposes qualitative empirical research, but the research is not completed at that time. (A later paper reports on this, with updated theoretical frameworks.) & 19 \\ \hline
Choosing the right prioritisation method \cite{hatton2008choosing} & 2008 & S. Hatton & An \textbf{experiment} with students was conducted to study requirements prioritization methods. & 31 \\ \hline
Key aspects of software release planning in industry \cite{lindgren2008key} & 2008 & M. Lindgren, R. Land, C. Norstrom, and A. Wall & \textbf{Case study} research is performed across multiple cases (organizations) to identify key aspects of release planning including several aspects involving time. & 19 \\ \hline
How do real options concepts fit in agile requirements engineering? \cite{racheva2010real} & 2010 & Z. Racheva and M. Daneva & A \textbf{hybrid research design} “combines a scoping review {[}6{]}, the CHAPL framework {[}7{]}, and a case study {[}8{]}.” In another place, it is described as a “cross-case study in eight organizations”, and it includes 11 interviews. & 3 \\ \hline
Software for scientists facing wicked problems lessons from the VISTAS Project \cite{cushing2015software} & 2015 & J. B. Cushing, K. M. Winters, and D. Lach & \textbf{Case study} research is conducted on a complex software project. The results include conclusions about the opportunities and challenges of embracing the complexities of wicked problems in such multi-stakeholder environments. & 1 \\ \hline
Managing technical debt in software projects using scrum: An action research \cite{oliveira2015managing} & 2015 & F. Oliveira, A. Goldman, and V. Santos & \textbf{Action Research}: A technical debt management framework is evaluated in the real context of software projects. & 2 \\ \hline
An empirically developed method to aid decisions on architectural Technical Debt Refactoring: AnaConDebt \cite{martini2016empirically} & 2016 & A. Martini and J. Bosch & \textbf{Design Science} Research evaluated in multiple \textbf{cases}. A method is developed iteratively with industry partners and then evaluated in a separately completed case. & 0 \\ \hline
\end{tabular}\end{table*}

We will focus later on this set of 13 papers in detail.

\subsection{Which dimensions are considered in these studies?}

Of the 155 papers coded, 88 identified at least one dimension of trade-off. The majority of those, 54, are choices within one or two dimensions, for example within differing stakeholder goals or between costs and time. 

\begin{figure}
\caption{Number of studies with n trade-off dimensions}
\label{dim-hist}
\centering
\includegraphics[width=8.0cm]{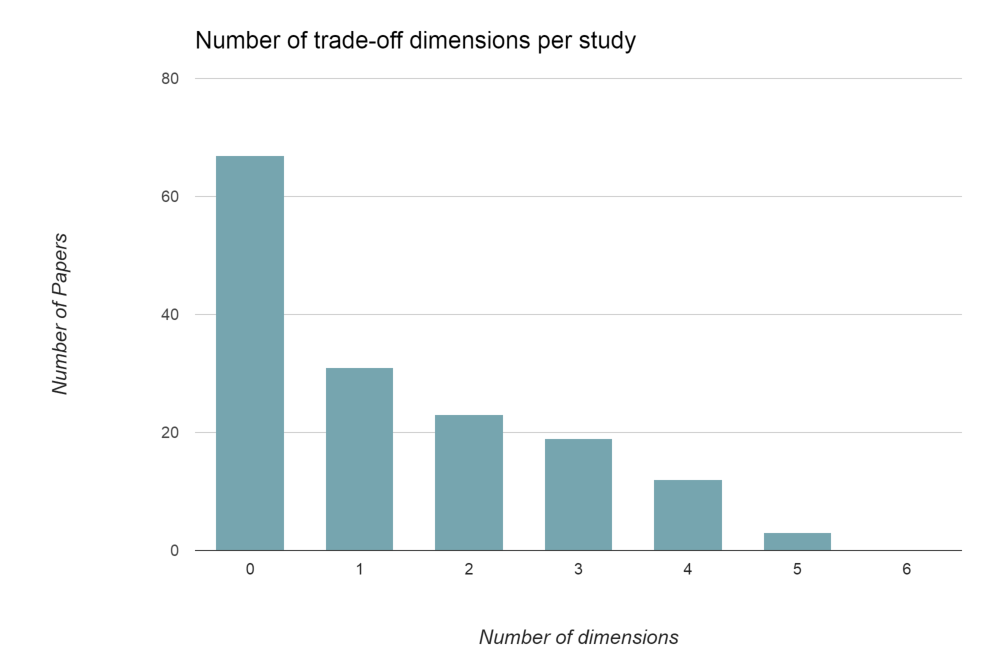}
\end{figure}

The most discussed single trade-off dimension is cost (39), the next highest mention is time (38), then quality (33), which includes nonfunctional requirements discussed as a group or specifically as ``usability,'' ``security,'' etc... The ``other'' category (38) had a diversity of dimensions seen in Table \ref{other-tradeoffs}, those that occurred more than 5 times are indicated with an asterisk. In a few cases we categorized terms as the same when the language had some variation (e.g. value, business value, and business benefits into ``value''). A record of these decisions was not recorded. 

\begin{table}[t]
\centering
\caption{Count of trade-off dimensions in all studies}
\label{all-tradeoffs}
\begin{tabular}{|ll|}
\hline
cost & 39 \\
time & 38 \\
other & 38 \\
quality & 33 \\
functionality & 16 \\
risk & 15 \\
value & 13 \\
maintenance & 5 \\ \hline
\end{tabular}
\end{table}

\begin{table}[]
\centering
\caption{``Other'' trade-off dimensions}
\label{other-tradeoffs}
\begin{tabular}{l}
\hline
\multicolumn{1}{|l|}{\begin{tabular}[c]{@{}l@{}}benefit*\\ competition\\ complexity\\ labour*\\ methodology*\\ opportunities\\ return on investment\\ goals*\\ technical debt\\ vendor\\ other\end{tabular}} \\ \hline
*those with an asterisk appeared more than 5 times
\end{tabular}
\end{table}

Of the 13 papers that empirically discussed a trade-off decision, the emphasis within dimensions was on cost (10), then time (7). Table \ref{emp-tradeoffs} shows the number of papers in which each dimension occurs within this set.

\begin{table}[]
\centering
\caption{Count of trade-off dimensions in 13 empirical studies}
\label{emp-tradeoffs}
\begin{tabular}{|ll|}
\hline
cost & 10 \\
time & 7 \\
quality & 6 \\
value & 4 \\
other (benefit) & 3 \\
other (goals) & 3 \\
functionality & 1 \\
other & 1 \\
other (competition) & 1 \\
other (methodology) & 1 \\ \hline
\end{tabular}
\end{table}

Further analysis of these dimensions could identify which sets of dimensions frequently co-occur. The data set has been prepared to support this analysis.

\subsection{How has the role of time been conceptualized in these studies?}

Time is the most popular dimension across all papers, 38 papers with time as a dimension; this is  unsurprising considering the search term. Within this, time is addressed in the various mapping groups: empirical (7), empirical component (19), non-empirical (12), and literature review (0).

However, the role of time is of course not always intertemporal. Time surfaces

\begin{itemize}
\item As the object of effort estimation: How much time will each of these options take?
\item As a factor in project management: How much time is available to the team?
\item As an attribute of decision making - how much time does it take to make a choice?
\item As a factor in scheduling, finally, the closest to intertemporal choice: Should we release now or later?
\end{itemize}

In order to characterize the role of time in the 13 empirical studies, the primary researcher performed a detailed analysis of the 13 papers that were coded as empirical studies of trade-off decisions involving time. Reading through the papers, the researcher performed iterative qualitative coding. He identified all mentioning of `time', reviewed each of their contexts, and iteratively developed a set of codes that described each new occurrence while continuing to apply to the existing occurrences. The resulting set of categories is summarized in Table \ref{time-usage}. Time in this set of papers is discussed most importantly as:

\begin{itemize}
\item A constrained resource in software project,
\item The time it takes to apply a method (e.g. designed by the researcher) in a project, and often a measure of that time,
\item an axis of discrete units `of time' over which a sequence of events take place, 
\item the time to market or the time to delivery, and
\item as an axis of change on which to pick suitable moments for action.
\end{itemize}

Additionally, a number of unique attributes and aspects surfaced once that were interpreted as tangential, since the concept of time was not central to the focus or nature of decision making. For example, this included a case that discussed technical consideration of real-time systems in project decisions or a discussion of how fixed-time release cycles provided consistent structure and rhythm to an organization's processes.

\begin{table*}[t]
	\centering
	\caption{Usage of time as a dimension}
	\label{time-usage}
	\begin{tabular}{|p{2.4cm}|p{1.5cm}|p{1.3cm}|p{1.8cm}|p{1.5cm}|p{2cm}|p{4.6cm}|}
		\hline
Time as...\newline \newline ... in & Constrained \newline Resource &  Time \newline to Apply & Sequence of \newline Project Events & Time \newline to Market & Axis of Change & Other: Time as... \\ \hline
		Abdel-Hamid \cite{abdel1999impact} & X &  &  &  &  &  \\ \hline
		Al-Naeem \cite{al2005quality} & X & X &  &  &  & ... in real-time systems \\ \hline
		Cushing \cite{cushing2015software} &  & X &  &  &  & ... a domain concern \\ \hline
		Filho \cite{simao2016task} & X &  &  &  &  & ... time zone differences in distributed teams \\ \hline
		Hatton \cite{hatton2008choosing} & X & X & X &  &  &  \\ \hline
		Karlsson \cite{karlsson1997cost} &  &  &  & X &  &  \\ \hline
		Lindgren \cite{lindgren2008key} & X &  &  & X &  Release Planning & ... rhythm of an organization \\ \hline
		Martini \cite{martini2016empirically} & X & X & X &  & Tech Debt &  \\ \hline
		Oliveira \cite{oliveira2015managing} & X & X & X &  &  &  \\ \hline
		Racheva \cite{racheva2010real} & X &  & X &  & Tech Debt &  \\ \hline
		Slaughter \cite{slaughter1998evaluating}& X &  & X &  &  &  \\ \hline
		Van Solingen \cite{van2004measuring} & X &  &  & X &  &  \\ \hline
		Wohlin \cite{wohlin2005important} & X & X &  & X &  Release Planning& ... path to an uncertain future of changing practice \\ \hline
		Zannier \cite{zannier2005qualitative} &  & X &  &  &  &  \\ \hline
	\end{tabular}
\end{table*}

It is the last conception listed above, the axis of change, where intertemporal decisions arise explicitly. In this set, they arose in particular in two specific forms, each represented by two distinct papers:

\begin{enumerate}
\item \textbf{Technical Debt} management raises questions around when to repay, and papers discussed how these decisions are being made;
\item \textbf{Release planning} raises questions of timing and of which \textbf{requirements to prioritize} and include for a given release.
\end{enumerate}

However, in neither of these cases was explicit attention given to behavioral insights, or into how decision makers arrive at their choices.

\subsection{Which assumptions on decision making underpin these studies?}

The predominant model of decision making in the relevant papers, so dominant that it is normally not made explicit, is a normative decision making model that builds on a Taylorist perspective on management, focused on efficiency and effectiveness, measured in the most scientifically accurate manner possible. Decision making assumes the presence and validity of normative theories of decision analysis in which clearly defined options are weighed against stated criteria to determine the best choice. The actual choosing is then often presumed to be unproblematic. Rational choice is the standard model, sometimes with explicit awareness of its limitations, often articulated in the frame of, or consistent with, bounded rationality. Yet, awareness is also present that ``research has proven that humans make trade-off analyses continuously-- if not on the basis of objective measurements then on intuition.'' (Van Solingen \cite{van2004measuring} pointing to Beach's Image Theory ). One paper explicitly proposes to contrast `rational' decision making with `naturalistic' frameworks and categorizes frameworks including Simon's Bounded Rationality, Prospect Theory, Image Theory, and other models \citeA{zannier2005qualitative}. The paper itself does not conduct empirical work, but a subsequent paper of the same authors does \cite{zannier2007model}.

In this paper, Zannier and Maurer conduct multiple interviews to develop an understanding of two modes of decision making characterized as rational and naturalistic \cite{zannier2007model}. The findings suggest a distinction between problem structuring and problem solving, and the authors conclude that both modes are relevant and one of them is typically the dominant approach. When the focus of decision making was on structuring, as was most commonly the case in the software design activities studied, naturalistic decision making modes dominated. Where the focus was on problem solving, rational modes dominated. In each modes, aspects of the other were present as well. The authors add that ``software designers often use satisficing and singular evaluation in trying different approaches to design'' -- where one option is checked for plausibility rather than being evaluated against other options, as described by  naturalistic decision making frameworks \cite{klein1999sources}.

However, the normative, rational decision making model also dominates the empirical papers that addressed the notion of trade-offs across time more explicitly.

In Lindgren's study of release planning, intertemporal considerations are foregrounded explicitly in discussions of short- and long-term planning, and an explicit connection is made to the need to repay technical debt \citeA{lindgren2008key}. However, while the paper highlights the need for longer-term perspectives and reports on empirical work, it focuses on larger questions and leaves open how precisely the decision makers acted in these decisions.

In Martini's study of technical debt management, time is similarly prominent: ``the TD theoretical framework instantiates a relationship between the cost and the impact of a single sub-optimal solution over time. In particular, the metaphor stresses the short-term gain given by a sub-optimal solution against the long-term one considered optimal'' \citeA{martini2016empirically}. Decisions have to be taken at the right time and have to anticipate uncertain future outcomes. The assumptions that are surfaced about the decision makers suggest that in the presence of perfect information, they would take the correct, optimal decision, surfacing assumptions of rational choice, contingent upon and bounded by the availability of information.

Racheva's study examines requirements prioritization in an agile environment at inter-iteration time-- the moment ``when requirements are re-prioritized in the face of project uncertainties''. At that stage, trade-offs consist in choosing what to do now and what not to do at the next iteration, a decision taken between iterations when requirements are re-prioritized \citeA{racheva2010real}. Over time, more information will be delivered and less of the limited resource of time will be available to act on it: ``The client can wait to the last responsible moment \dots to make his decision\dots The term `responsible' means that the client needs to understand the last point of time to make a decision without affecting the delivery of the project'' \citeA{racheva2010real}. The framework that is introduced aims to provide a conceptual frame for uncertainty over time by introducing Real Options Analysis. Given the agile focus, it is unsurprising that the research focuses on grounding proposals on empirical insights and being responsive to the actual behavior of practitioners. The underpinning assumptions are implicit, but build clearly on ideas of bounded rationality.

Finally, Wohlin's paper on release planning aims to provide guidance for which types of criteria practitioners should consider when conducting requirements prioritization for release planning. However, the actual decision making is not discussed \citeA{wohlin2005important}.

\begin{table}
\centering
\caption{Number of papers with time as a trade-off dimension}
\label{time-count}
\begin{tabular}{|l|l|}
\hline
Empirical           & 7  \\ \hline
Empirical Component & 19 \\ \hline
Non Empirical       & 12 \\ \hline
Literature Review   & 0  \\ \hline
Total               & 38 \\ \hline
\end{tabular}
\end{table}

\section{Conclusions and Outlook}

\subsection{Discussion}

Trade-off decisions in software engineering have been studied and modeled for a long time. However, the role of time in most studies focuses on time as a limited resource and a ticking clock. However, some work, such as technical debt research, has foregrounded the attention to trade-off decisions across time.

In general, it is difficult to find the most relevant work on such a subtle topic as `decision making involving trade-off decisions across time', because the terminology that can be used to describe it is not stabilized and thus the terms are used in ambiguous and varied ways. This means that we cannot assume we have covered the body of literature that in fact discusses these questions comprehensively. However, because the main goal is to understand common assumptions and norms, a comprehensive identification of all works having studied these aspects in depth is not the primary criterion.

A number of studies have suggested that normative models are inadequate in explaining how people actually take decisions \cite{frederick2002time}. Behavioral perspectives and empirical research are needed to provide new and deeper understanding on the practice of software engineering \cite{lenberg2015behavioral}. This suggests that more descriptive research is needed to provide a bottom-up empirically grounded description of decision making. We are aware of some studies within the domain of SE, but none focused on time trade-offs.

There is awareness in parts of the empirical literature that normative decision theory has limited relevance for descriptive and explanatory purposes. Nonetheless, little empirical work surfaced that explicitly pursues empirically grounded, descriptive approaches, and none that studied trade-offs in time in depth. 

However, the most explicit discussion of intertemporal choice and trade-offs were found in decisions about technical debt management. This suggests that the body of work on technical debt should be analyzed in more depth to characterize the tension between normative decision theory and descriptive approaches and to identify opportunities to improve our understanding of trade-off decision making in practice.

\subsection{Threats to Validity}

\subsubsection{Internal Validity}

Although we followed Kitchenham's procedure for systematic literature reviews, minor deviations from the protocol should be noted:

\begin{itemize}
\item The documented of detailed codes being merged is not comprehensive. These codes were merged carefully and only when the terms were close, as discussed; however, this limits the traceability of analysis.
\item Within the TD set, a corpus was constructed to enable future corpus-assisted discourse analysis. However, we were unable to include 4 papers which could not be converted from .pdf to text (presumably due to their encoding).
\end{itemize}

\subsubsection{External Validity}

The searches were limited to 3 databases, and no snowballing was conducted. This limits the external validity of our findings. However, the databases we used are commonly considered the main sources, and Google Scholar is often seen as the `most comprehensive' source.

By including ``time'' in our search term, we wanted to get a sense of how time was treated as a dimension in empirical discussions of trade-offs in SE. However, not all relevant papers discuss time in this manner: some are about SE decision making in general. We were not trying to examine the assumptions underlying SE decision making in general, and our results cannot be generalized as such. 

Some of the test searches also included the term ``requirements engineering'', but this term was later dropped. While it is plausible that many of the results could be captured by software engineering, the search cannot be said to be generalizable to requirements engineering.

\subsubsection{Construct Validity}

Intertemporal choice is not a term that is used in SE literature. Indeed, there appears to be no blanket term for describing the types of decisions and tradeoffs that we attempt to study in this literature review. 
We thus refrained from the use of terms specific to the domain of intertemporal choice to ensure we identify how the SE community talks about these concerns. We believe this is an adequate measure to tease out intertemporal choice from the larger body of SE literature, but this cannot be guaranteed, as authors may speak about intertemporal choice in different terms that may  have been missed by our search.

The search terms are known to be incomplete in the sense that terms related to `time' and `decision making', and disciplinary terms such as ``requirements engineering'', have not been included in the search. This choice was taken since the aim was to identify common assumptions and decision making theories within an acceptable time frame.

\subsubsection{Reliability}

Not all researchers were trained software engineers, constituting a threat to the reliability of the literature review because terms might not be properly understood. This had to be addressed and considered in the setup of the protocol, as discussed above: We conducted iterative coding, reviewed with the internal expert, and erred on the side of caution. In order to ensure reliability in relevancy rating and coding, researchers conducted the work individually and then afterward compared. Where there was discontinuity between voters or where they were not sure how to evaluate a paper, results were obtained by consensus. If errors or ambiguities in the review process were discovered, another iteration was done to correct for that. 

The internal and external review were conducted from a perspective of software engineering expertise.

\subsection{Future work}

To prepare the empirical study of time trade-offs, an analysis of the technical debt corpus is the logical next step: This work is very clearly bounded and focused on a closely related area of high relevance for the SE community. As part of this literature review, we have prepared a text corpus with the identified 231\footnote{Down from 246.} papers that can be downloaded and analyzed quantitatively. This will make it possible to use complementary techniques such as corpus-assisted discourse analysis to identify the associations and meanings attributed to time trade-off decisions in the domain of technical debt management.

\appendices
\section{Pretest Search Strings}
\label{pretest-strings}

\subsection{Pretest 1: June 30, 2016}

\begin{itemize}
\item tradeoff \textbf{AND} software engineering \textbf{OR} requirement* engineering
\item intertemporal choice \textbf{AND} software engineering \textbf{OR} requirement* engineering
\item behavioral economic* \textbf{AND} software engineering \textbf{OR} requirement* engineering
\end{itemize}

\subsection{Pretest 2: October 28, 2016}
\label{long-strings}
\begin{itemize}
\item trade-off \textbf{OR} tradeoff \textbf{OR} "trade off" \textbf{OR} conflict \\
\textbf{AND} long-living \textbf{OR} "long living" \textbf{OR} "long lasting" \textbf{OR} "long-lasting" \textbf{OR} longevity \textbf{OR} "long term" \textbf{OR} "end of life" \textbf{OR} end-of-life \textbf{OR} future \textbf{OR} "life cycle" \textbf{OR} "life-cycle" \textbf{OR} "lifecycle" \textbf{OR} enduring \textbf{OR} temporal \textbf{OR} sustain* \\
\textbf{AND} "requirement engineering" \textbf{OR} "requirements engineering" \\
\item (trade-off \textbf{OR} tradeoff \textbf{OR} "trade off" \textbf{OR} conflict) \\
\textbf{AND} (long-living \textbf{OR} "long living" \textbf{OR} "long lasting" \textbf{OR} "long-lasting" \textbf{OR} longevity \textbf{OR} "long term" \textbf{OR} "end of life" \textbf{OR} “end-of-life” \textbf{OR} future \textbf{OR} "life cycle" \textbf{OR} "life-cycle" \textbf{OR} "lifecycle" \textbf{OR} enduring \textbf{OR} temporal \textbf{OR} sustain*) \\
\textbf{AND} (stakeholder \textbf{OR} values \textbf{OR} preferences \textbf{OR} goals \textbf{OR} benefits \textbf{OR} elicitation \textbf{OR} negotiation \textbf{OR} prioritization \textbf{OR} incentive)
\textbf{AND} ("software engineering")
\end{itemize}

\section{Trade-off Dimensions Descriptions}
\label{trade-off-descriptions}
Here we captured self-reported dimensions that we encountered while coding: 

\begin{itemize}
\item (benefit) as a direct component of cost-benefit analysis
\item (competition)
\item (complexity) – Complexity of project architecture/code
\item (labour)
\item (methodology) – software development methodology or system development life cycle 
\item (opportunities)
\item (ROI) – Return on Investment
\item (goals) – goals
\item (techdebt) – Technical Debt
\item (vendor) – vendor viability
\end{itemize}

\section{Selected Papers}
\label{selected-papers}
\renewcommand{\refname}{Research has an Empirical Component which Studies Decisions}

\renewcommand{\refname}{Research has an Empirical Component \& Discusses Trade-offs or Like Decisions}

\renewcommand{\refname}{Research has an Empirical Component \& does not Discuss Trade-offs or Like Decisions}

\renewcommand{\refname}{Research does not have an Empirical component \& Discusses Trade-offs or like Decisions}

\renewcommand{\refname}{Research does not have an Empirical Component\& does not Discuss Trade-offs or Like Decisions}

\section*{Acknowledgment}

Part of this work was supported by NSERC through RGPIN-2016-06640, and by the Connaught Fund. Special thanks to Ruzanna Chitchyan and Stefanie Betz for reviewing the literature review protocol.

\ifCLASSOPTIONcaptionsoff
  \newpage
\fi



\renewcommand{\refname}{References}


\begin{thebibliography}{10}
\providecommand{\url}[1]{#1}
\csname url@samestyle\endcsname
\providecommand{\newblock}{\relax}
\providecommand{\bibinfo}[2]{#2}
\providecommand{\BIBentrySTDinterwordspacing}{\spaceskip=0pt\relax}
\providecommand{\BIBentryALTinterwordstretchfactor}{4}
\providecommand{\BIBentryALTinterwordspacing}{\spaceskip=\fontdimen2\font plus
\BIBentryALTinterwordstretchfactor\fontdimen3\font minus
  \fontdimen4\font\relax}
\providecommand{\BIBforeignlanguage}[2]{{%
\expandafter\ifx\csname l@#1\endcsname\relax
\typeout{** WARNING: IEEEtran.bst: No hyphenation pattern has been}%
\typeout{** loaded for the language `#1'. Using the pattern for}%
\typeout{** the default language instead.}%
\else
\language=\csname l@#1\endcsname
\fi
#2}}
\providecommand{\BIBdecl}{\relax}
\BIBdecl

\bibitem[A1]{zannier2005qualitative}
C.~Zannier and F.~Maurer, ``A qualitative empirical evaluation of design
  decisions,'' in \emph{ACM SIGSOFT Software Engineering Notes}, vol.~30,
  no.~4.\hskip 1em plus 0.5em minus 0.4em\relax ACM, 2005, pp. 1--7.

\bibitem[A2]{lindgren2008key}
M.~Lindgren, R.~Land, C.~Norstr{\"o}m, and A.~Wall, ``Key aspects of software
  release planning in industry,'' in \emph{Software Engineering, 2008. ASWEC
  2008. 19th Australian Conference on}.\hskip 1em plus 0.5em minus 0.4em\relax
  IEEE, 2008, pp. 320--329.

\bibitem[A3]{martini2016empirically}
A.~Martini and J.~Bosch, ``An empirically developed method to aid decisions on
  architectural technical debt refactoring: Anacondebt,'' in \emph{Proceedings
  of the 38th International Conference on Software Engineering
  Companion}.\hskip 1em plus 0.5em minus 0.4em\relax ACM, 2016, pp. 31--40.

\bibitem[A4]{racheva2010real}
Z.~Racheva and M.~Daneva, ``How do real options concepts fit in agile
  requirements engineering?'' in \emph{Software Engineering Research,
  Management and Applications (SERA), 2010 Eighth ACIS International Conference
  on}.\hskip 1em plus 0.5em minus 0.4em\relax IEEE, 2010, pp. 231--238.

\bibitem[A5]{wohlin2005important}
C.~Wohlin and A.~Aurum, ``What is important when deciding to include a software
  requirement in a project or release?'' in \emph{Empirical Software
  Engineering, 2005. 2005 International Symposium on}.\hskip 1em plus 0.5em
  minus 0.4em\relax IEEE, 2005, pp. 10--pp.

\bibitem[A6]{abdel1999impact}
T.~K. Abdel-Hamid, K.~Sengupta, and C.~Swett, ``The impact of goals on software
  project management: An experimental investigation,'' \emph{MIS quarterly},
  pp. 531--555, 1999.

\bibitem[A7]{al2005quality}
T.~Al-Naeem, I.~Gorton, M.~A. Babar, F.~Rabhi, and B.~Benatallah, ``A
  quality-driven systematic approach for architecting distributed software
  applications,'' in \emph{Proceedings of the 27th international conference on
  Software engineering}.\hskip 1em plus 0.5em minus 0.4em\relax ACM, 2005, pp.
  244--253.

\bibitem[A8]{cushing2015software}
J.~B. Cushing, K.~M. Winters, and D.~Lach, ``Software for scientists facing
  wicked problems lessons from the vistas project,'' in \emph{Proceedings of
  the 16th Annual International Conference on Digital Government
  Research}.\hskip 1em plus 0.5em minus 0.4em\relax ACM, 2015, pp. 61--70.

\bibitem[A9]{hatton2008choosing}
S.~Hatton, ``Choosing the right prioritisation method,'' in \emph{Software
  Engineering, 2008. ASWEC 2008. 19th Australian Conference on}.\hskip 1em plus
  0.5em minus 0.4em\relax IEEE, 2008, pp. 517--526.

\bibitem[A10]{karlsson1997cost}
J.~Karlsson and K.~Ryan, ``A cost-value approach for prioritizing
  requirements,'' \emph{IEEE software}, vol.~14, no.~5, pp. 67--74, 1997.

\bibitem[A11]{oliveira2015managing}
F.~Oliveira, A.~Goldman, and V.~Santos, ``Managing technical debt in software
  projects using scrum: An action research,'' in \emph{Agile Conference
  (AGILE), 2015}.\hskip 1em plus 0.5em minus 0.4em\relax IEEE, 2015, pp.
  50--59.

\bibitem[A12]{slaughter1998evaluating}
S.~A. Slaughter, D.~E. Harter, and M.~S. Krishnan, ``Evaluating the cost of
  software quality,'' \emph{Communications of the ACM}, vol.~41, no.~8, pp.
  67--73, 1998.

\bibitem[A13]{van2004measuring}
R.~Van~Solingen, ``Measuring the roi of software process improvement,''
  \emph{IEEE software}, vol.~21, no.~3, pp. 32--38, 2004.

\end{thebibliography}

\begin{thebibliography}{10}
\providecommand{\url}[1]{#1}
\csname url@samestyle\endcsname
\providecommand{\newblock}{\relax}
\providecommand{\bibinfo}[2]{#2}
\providecommand{\BIBentrySTDinterwordspacing}{\spaceskip=0pt\relax}
\providecommand{\BIBentryALTinterwordstretchfactor}{4}
\providecommand{\BIBentryALTinterwordspacing}{\spaceskip=\fontdimen2\font plus
\BIBentryALTinterwordstretchfactor\fontdimen3\font minus
  \fontdimen4\font\relax}
\providecommand{\BIBforeignlanguage}[2]{{%
\expandafter\ifx\csname l@#1\endcsname\relax
\typeout{** WARNING: IEEEtran.bst: No hyphenation pattern has been}%
\typeout{** loaded for the language `#1'. Using the pattern for}%
\typeout{** the default language instead.}%
\else
\language=\csname l@#1\endcsname
\fi
#2}}
\providecommand{\BIBdecl}{\relax}
\BIBdecl

\bibitem[B1]{simao2016task}
M.~Sim{\~a}o~Filho, P.~R. Pinheiro, and A.~B. Albuquerque, ``Task allocation in
  distributed software development aided by verbal decision analysis,'' in
  \emph{Software Engineering Perspectives and Application in Intelligent
  Systems}.\hskip 1em plus 0.5em minus 0.4em\relax Springer, 2016, pp.
  127--137.

\bibitem[B2]{aguilar2001evolutionary}
J.~S. Aguilar-Ruiz, I.~Ramos, J.~C. Riquelme, and M.~Toro, ``An evolutionary
  approach to estimating software development projects,'' \emph{Information and
  Software Technology}, vol.~43, no.~14, pp. 875--882, 2001.

\bibitem[B3]{azar2007value}
J.~Azar, R.~K. Smith, and D.~Cordes, ``Value-oriented requirements
  prioritization in a small development organization,'' \emph{IEEE software},
  vol.~24, no.~1, 2007.

\bibitem[B4]{buyukozkan2004fuzzy}
G.~B{\"u}y{\"u}k{\"o}zkan and O.~Feyz{\i}og̃lu, ``A fuzzy-logic-based
  decision-making approach for new product development,'' \emph{International
  journal of production economics}, vol.~90, no.~1, pp. 27--45, 2004.

\bibitem[B5]{bajaj2013multi}
P.~Bajaj and V.~Arora, ``Multi-person decision-making for requirements
  prioritization using fuzzy ahp,'' \emph{ACM SIGSOFT Software Engineering
  Notes}, vol.~38, no.~5, pp. 1--6, 2013.

\bibitem[B6]{bano2014users}
M.~Bano and D.~Zowghi, ``Users' voice and service selection: An empirical
  study,'' in \emph{Empirical Requirements Engineering (EmpiRE), 2014 IEEE
  Fourth International Workshop on}.\hskip 1em plus 0.5em minus 0.4em\relax
  IEEE, 2014, pp. 76--79.

\bibitem[B7]{bohnet2011monitoring}
J.~Bohnet and J.~D{\"o}llner, ``Monitoring code quality and development
  activity by software maps,'' in \emph{Proceedings of the 2nd Workshop on
  Managing Technical Debt}.\hskip 1em plus 0.5em minus 0.4em\relax ACM, 2011,
  pp. 9--16.

\bibitem[B8]{boness2002data}
K.~Boness and R.~Harrison, ``A data collection case study supporting
  requirements oriented prediction and management in software developments,''
  in \emph{Computer Software and Applications Conference, 2002. COMPSAC 2002.
  Proceedings. 26th Annual International}.\hskip 1em plus 0.5em minus
  0.4em\relax IEEE, 2002, pp. 744--746.

\bibitem[B9]{choetkiertikul2015predicting}
M.~Choetkiertikul, H.~K. Dam, T.~Tran, and A.~Ghose, ``Predicting delays in
  software projects using networked classification (t),'' in \emph{Automated
  Software Engineering (ASE), 2015 30th IEEE/ACM International Conference
  on}.\hskip 1em plus 0.5em minus 0.4em\relax IEEE, 2015, pp. 353--364.

\bibitem[B10]{dalcher2005development}
D.~Dalcher, O.~Benediktsson, and H.~Thorbergsson, ``Development life cycle
  management: a multiproject experiment,'' in \emph{Engineering of
  Computer-Based Systems, 2005. ECBS'05. 12th IEEE International Conference and
  Workshops on the}.\hskip 1em plus 0.5em minus 0.4em\relax IEEE, 2005, pp.
  289--296.

\bibitem[B11]{dong2011value}
X.~Dong, Q.-S. Yang, Q.~Wang, J.~Zhai, and G.~Ruhe, ``Value-risk trade-off
  analysis for iteration planning in extreme programming,'' in \emph{Software
  Engineering Conference (APSEC), 2011 18th Asia Pacific}.\hskip 1em plus 0.5em
  minus 0.4em\relax IEEE, 2011, pp. 397--404.

\bibitem[B12]{ebert1999technical}
C.~Ebert, ``Technical controlling in software development,''
  \emph{International Journal of Project Management}, vol.~17, no.~1, pp.
  17--28, 1999.

\bibitem[B13]{egyed2006support}
A.~Egyed and D.~S. Wile, ``Support for managing design-time decisions,''
  \emph{IEEE Transactions on Software Engineering}, vol.~32, no.~5, pp.
  299--314, 2006.

\bibitem[B14]{erofeev2006usage}
S.~Erofeev and P.~Di~Giacomo, ``Usage of dynamic decision models as an agile
  approach to cots taxonomies construction,'' in \emph{Commercial-off-the-Shelf
  (COTS)-Based Software Systems, 2006. Fifth International Conference
  on}.\hskip 1em plus 0.5em minus 0.4em\relax IEEE, 2006, pp. 8--pp.

\bibitem[B15]{falessi2013practical}
D.~Falessi, M.~A. Shaw, F.~Shull, K.~Mullen, and M.~S. Keymind, ``Practical
  considerations, challenges, and requirements of tool-support for managing
  technical debt,'' in \emph{Managing Technical Debt (MTD), 2013 4th
  International Workshop on}.\hskip 1em plus 0.5em minus 0.4em\relax IEEE,
  2013, pp. 16--19.

\bibitem[B16]{garg2011stop}
M.~Garg, R.~Lai, and S.~J. Huang, ``When to stop testing: a study from the
  perspective of software reliability models,'' \emph{IET software}, vol.~5,
  no.~3, pp. 263--273, 2011.

\bibitem[B17]{gaur2010application}
V.~Gaur, A.~Soni, and P.~Bedi, ``An application of multi-person decision-making
  model for negotiating and prioritizing requirements in agent-oriented
  paradigm,'' in \emph{Data Storage and Data Engineering (DSDE), 2010
  International Conference on}.\hskip 1em plus 0.5em minus 0.4em\relax IEEE,
  2010, pp. 164--168.

\bibitem[B18]{hassan2015minimizing}
S.~Hassan, N.~Bencomo, and R.~Bahsoon, ``Minimizing nasty surprises with better
  informed decision-making in self-adaptive systems,'' in \emph{Proceedings of
  the 10th International Symposium on Software Engineering for Adaptive and
  Self-Managing Systems}.\hskip 1em plus 0.5em minus 0.4em\relax IEEE Press,
  2015, pp. 134--144.

\bibitem[B19]{hedberg2009integrating}
H.~Hedberg and N.~Iivari, ``Integrating hci specialists into open source
  software development projects,'' in \emph{IFIP International Conference on
  Open Source Systems}.\hskip 1em plus 0.5em minus 0.4em\relax Springer, 2009,
  pp. 251--263.

\bibitem[B20]{jedlitschka2007relevant}
A.~Jedlitschka, M.~Ciolkowski, C.~Denger, B.~Freimut, and A.~Schlichting,
  ``Relevant information sources for successful technology transfer: a survey
  using inspections as an example,'' in \emph{Empirical Software Engineering
  and Measurement, 2007. ESEM 2007. First International Symposium on}.\hskip
  1em plus 0.5em minus 0.4em\relax IEEE, 2007, pp. 31--40.

\bibitem[B21]{khan2014method}
M.~A. Khan, A.~Parveen, and M.~Sadiq, ``A method for the selection of software
  development life cycle models using analytic hierarchy process,'' in
  \emph{Issues and Challenges in Intelligent Computing Techniques (ICICT), 2014
  International Conference on}.\hskip 1em plus 0.5em minus 0.4em\relax IEEE,
  2014, pp. 534--540.

\bibitem[B22]{khan2016repizer}
S.~U.~R. Khan, S.~P. Lee, M.~Dabbagh, M.~Tahir, M.~Khan, and M.~Arif,
  ``Repizer: a framework for prioritization of software requirements,''
  \emph{Frontiers of Information Technology \& Electronic Engineering},
  vol.~17, pp. 750--765, 2016.

\bibitem[B23]{Khriss2000}
I.~Khriss, R.~K. Keller, and I.~A. Hamid, ``Pattern-based refinement schemas
  for design knowledge transfer,'' \emph{Knowledge-Based Systems}, vol.~13,
  no.~6, pp. 403--415, 2000.

\bibitem[B24]{kim2007lightweight}
C.-K. Kim, D.-H. Lee, I.-Y. Ko, and J.~Baik, ``A lightweight value-based
  software architecture evaluation,'' in \emph{Software Engineering, Artificial
  Intelligence, Networking, and Parallel/Distributed Computing, 2007. SNPD
  2007. Eighth ACIS International Conference on}, vol.~2.\hskip 1em plus 0.5em
  minus 0.4em\relax IEEE, 2007, pp. 646--649.

\bibitem[B25]{kimura2009new}
M.~Kimura and T.~Fujiwara, ``A new criterion for the optimal software release
  problems: Moving average quality control chart with bootstrap sampling,'' in
  \emph{International Conference on Advanced Software Engineering and Its
  Applications}.\hskip 1em plus 0.5em minus 0.4em\relax Springer, 2009, pp.
  280--287.

\bibitem[B26]{krishnan2004decision}
M.~S. Krishnan, T.~Mukhopadhyay, and C.~H. Kriebel, ``A decision model for
  software maintenance,'' \emph{Information Systems Research}, vol.~15, no.~4,
  pp. 396--412, 2004.

\bibitem[B27]{kumar2013enterprise}
N.~Kumar and K.~Srivathsan, ``Enterprise risk evaluation and continuous
  mitigation using the fuzzy-multiattribute decision making-a conceptual
  approach,'' \emph{Journal of the Indian Institute of Science}, vol.~86,
  no.~6, p. 625, 2013.

\bibitem[B28]{machado2015project}
T.~C.~S. Machado, P.~R. Pinheiro, and I.~Tamanini, ``Project management aided
  by verbal decision analysis approaches: a case study for the selection of the
  best scrum practices,'' \emph{International Transactions in Operational
  Research}, vol.~22, no.~2, pp. 287--312, 2015.

\bibitem[B29]{neves2013context}
R.~Neves-Silva \emph{et~al.}, ``Context sensitive solution for collaborative
  decision making on quality assurance in software development processes,'' in
  \emph{Intelligent Decision Technologies: Proceedings of the 5th KES
  International Conference on Intelligent Decision Technologies (KES-IDT
  2013)}, vol. 255.\hskip 1em plus 0.5em minus 0.4em\relax IOS Press, 2013, p.
  130.

\bibitem[B30]{nguyen2014role}
A.~Nguyen-Duc, D.~S. Cruzes, and R.~Conradi, ``On the role of boundary spanners
  as team coordination mechanisms in organizationally distributed projects,''
  in \emph{Global Software Engineering (ICGSE), 2014 IEEE 9th International
  Conference on}.\hskip 1em plus 0.5em minus 0.4em\relax IEEE, 2014, pp.
  125--134.

\bibitem[B31]{nidhra2012analytical}
S.~Nidhra, L.~P.~K. Satish, and V.~S. Ethiraj, ``Analytical hierarchy process
  issues and mitigation strategy for large number of requirements,'' in
  \emph{Software Engineering (CONSEG), 2012 CSI Sixth International Conference
  on}.\hskip 1em plus 0.5em minus 0.4em\relax IEEE, 2012, pp. 1--8.

\bibitem[B32]{otero2010multi}
C.~E. Otero, L.~D. Otero, I.~Weissberger, and A.~Qureshi, ``A multi-criteria
  decision making approach for resource allocation in software engineering,''
  in \emph{Computer Modelling and Simulation (UKSim), 2010 12th International
  Conference on}.\hskip 1em plus 0.5em minus 0.4em\relax IEEE, 2010, pp.
  137--141.

\bibitem[B33]{plate2015impact}
H.~Plate, S.~E. Ponta, and A.~Sabetta, ``Impact assessment for vulnerabilities
  in open-source software libraries,'' in \emph{Software Maintenance and
  Evolution (ICSME), 2015 IEEE International Conference on}.\hskip 1em plus
  0.5em minus 0.4em\relax IEEE, 2015, pp. 411--420.

\bibitem[B34]{pozgaj2004strategies}
Z.~Pozgaj, H.~Sertic, and M.~Boban, ``Strategies for successful software
  development project preparation,'' in \emph{Information Technology
  Interfaces, 2004. 26th International Conference on}.\hskip 1em plus 0.5em
  minus 0.4em\relax IEEE, 2004, pp. 679--684.

\bibitem[B35]{racheva2010conceptual}
Z.~Racheva, M.~Daneva, A.~Herrmann, and R.~J. Wieringa, ``A conceptual model
  and process for client-driven agile requirements prioritization,'' in
  \emph{Research Challenges in Information Science (RCIS), 2010 Fourth
  International Conference on}.\hskip 1em plus 0.5em minus 0.4em\relax IEEE,
  2010, pp. 287--298.

\bibitem[B36]{raffo2002software}
D.~M. Raffo, W.~Harrison, and J.~Vandeville, ``Software process decision
  support: making process tradeoffs using a hybrid metrics, modeling and
  utility framework,'' in \emph{Proceedings of the 14th international
  conference on Software engineering and knowledge engineering}.\hskip 1em plus
  0.5em minus 0.4em\relax ACM, 2002, pp. 803--809.

\bibitem[B37]{ruhe2003trade}
G.~Ruhe, A.~Eberlein, and D.~Pfahl, ``Trade-off analysis for requirements
  selection,'' \emph{International Journal of Software Engineering and
  Knowledge Engineering}, vol.~13, no.~04, pp. 345--366, 2003.

\bibitem[B38]{sadiq2015method}
M.~Sadiq and S.~Sultana, ``A method for the selection of software testing
  techniques using analytic hierarchy process,'' in \emph{Computational
  Intelligence in Data Mining-Volume 1}.\hskip 1em plus 0.5em minus 0.4em\relax
  Springer, 2015, pp. 213--220.

\bibitem[B39]{sellier2008managing}
D.~Sellier, M.~Mannion, and J.~X. Mansell, ``Managing requirements
  inter-dependency for software product line derivation,'' \emph{Requirements
  engineering}, vol.~13, no.~4, pp. 299--313, 2008.

\bibitem[B40]{sengupta1999coping}
K.~Sengupta, T.~K. Abdel-Hamid, and M.~Bosley, ``Coping with staffing delays in
  software project management: an experimental investigation,'' \emph{IEEE
  Transactions on Systems, Man, and Cybernetics-Part A: Systems and Humans},
  vol.~29, no.~1, pp. 77--91, 1999.

\bibitem[B41]{sharma2014design}
S.~Sharma and S.~Srivastava, ``Design \& evaluation of hybrid framework for
  software maintenance,'' in \emph{Advance Computing Conference (IACC), 2014
  IEEE International}.\hskip 1em plus 0.5em minus 0.4em\relax IEEE, 2014, pp.
  1459--1468.

\bibitem[B42]{shepperd1989metrics}
M.~Shepperd, ``A metrics based tool for software design,'' in \emph{Software
  Engineering for Real Time Systems, 1989., Second International Conference
  on}.\hskip 1em plus 0.5em minus 0.4em\relax IET, 1989, pp. 45--49.

\bibitem[B43]{stoica2015system}
A.-J. Stoica, K.~Pelckmans, and W.~Rowe, ``System components of a general
  theory of software engineering,'' \emph{Science of Computer Programming},
  vol. 101, pp. 42--65, 2015.

\bibitem[B44]{wang2003fuzzy}
J.~Wang and Y.-I. Lin, ``A fuzzy multicriteria group decision making approach
  to select configuration items for software development,'' \emph{Fuzzy Sets
  and Systems}, vol. 134, no.~3, pp. 343--363, 2003.

\bibitem[B45]{wei2014implementation}
B.~Wei, Z.~Jin, D.~Zowghi, and B.~Yin, ``Implementation decision making for
  internetware driven by quality requirements,'' \emph{Science China
  Information Sciences}, vol.~57, no.~7, pp. 1--19, 2014.

\bibitem[B46]{welker1997development}
K.~D. Welker, P.~W. Oman, and G.~G. Atkinson, ``Development and application of
  an automated source code maintainability index,'' \emph{Journal of Software:
  Evolution and Process}, vol.~9, no.~3, pp. 127--159, 1997.

\end{thebibliography}

\begin{thebibliography}{10}
\providecommand{\url}[1]{#1}
\csname url@samestyle\endcsname
\providecommand{\newblock}{\relax}
\providecommand{\bibinfo}[2]{#2}
\providecommand{\BIBentrySTDinterwordspacing}{\spaceskip=0pt\relax}
\providecommand{\BIBentryALTinterwordstretchfactor}{4}
\providecommand{\BIBentryALTinterwordspacing}{\spaceskip=\fontdimen2\font plus
\BIBentryALTinterwordstretchfactor\fontdimen3\font minus
  \fontdimen4\font\relax}
\providecommand{\BIBforeignlanguage}[2]{{%
\expandafter\ifx\csname l@#1\endcsname\relax
\typeout{** WARNING: IEEEtran.bst: No hyphenation pattern has been}%
\typeout{** loaded for the language `#1'. Using the pattern for}%
\typeout{** the default language instead.}%
\else
\language=\csname l@#1\endcsname
\fi
#2}}
\providecommand{\BIBdecl}{\relax}
\BIBdecl

\bibitem[C1]{agarwal2006defining}
N.~Agarwal and U.~Rathod, ``Defining ÔsuccessÕ for software projects: An
  exploratory revelation,'' \emph{International journal of project management},
  vol.~24, no.~4, pp. 358--370, 2006.

\bibitem[C2]{badura2010coding}
V.~Badura, A.~Read, R.~O. Briggs, and G.-J. De~Vreede, ``Coding for unique
  ideas and ambiguity: Measuring the effects of a convergence intervention on
  the artifact of an ideation activity,'' in \emph{System Sciences (HICSS),
  2010 43rd Hawaii International Conference on}.\hskip 1em plus 0.5em minus
  0.4em\relax IEEE, 2010, pp. 1--10.

\bibitem[C3]{bernardo2009using}
C.~G. Bernardo, D.~A. Montini, D.~D. Fernandes, D.~A. da~Silva, L.~A.~V. Dias,
  and A.~M. da~Cunha, ``Using gqm for testing design patterns in real-time and
  embedded systems on a software production line,'' in \emph{Information
  Technology: New Generations, 2009. ITNG'09. Sixth International Conference
  on}.\hskip 1em plus 0.5em minus 0.4em\relax IEEE, 2009, pp. 1397--1404.

\bibitem[C4]{burns1987communications}
A.~Burns and M.~A. Rathwell, ``A communications environment for co-operative
  information systems development,'' \emph{Software Engineering Journal},
  vol.~2, no.~1, pp. 9--14, 1987.

\bibitem[C5]{buse2012information}
R.~P. Buse and T.~Zimmermann, ``Information needs for software development
  analytics,'' in \emph{Proceedings of the 34th international conference on
  software engineering}.\hskip 1em plus 0.5em minus 0.4em\relax IEEE Press,
  2012, pp. 987--996.

\bibitem[C6]{cottam2008representing}
J.~A. Cottam, J.~Hursey, and A.~Lumsdaine, ``Representing unit test data for
  large scale software development,'' in \emph{Proceedings of the 4th ACM
  symposium on Software visualization}.\hskip 1em plus 0.5em minus 0.4em\relax
  ACM, 2008, pp. 57--66.

\bibitem[C7]{dafoulas2001facilitating}
G.~A. Dafoulas and L.~A. Macaulay, ``Facilitating group formation and role
  allocation in software engineering groups,'' in \emph{Computer Systems and
  Applications, ACS/IEEE International Conference on. 2001}.\hskip 1em plus
  0.5em minus 0.4em\relax IEEE, 2001, pp. 352--359.

\bibitem[C8]{och2004speeding}
J.~N. och Dag, V.~Gervasi, S.~Brinkkemper, and B.~Regnell, ``Speeding up
  requirements management in a product software company: Linking customer
  wishes to product requirements through linguistic engineering,'' in
  \emph{Requirements Engineering Conference, 2004. Proceedings. 12th IEEE
  International}.\hskip 1em plus 0.5em minus 0.4em\relax IEEE, 2004, pp.
  283--294.

\bibitem[C9]{de2003silver}
G.-J. De~Vreede, R.~M. Davison, and R.~O. Briggs, ``How a silver bullet may
  lose its shine,'' \emph{Communications of the ACM}, vol.~46, no.~8, pp.
  96--101, 2003.

\bibitem[C10]{ekanayake2009tracking}
J.~Ekanayake, J.~Tappolet, H.~C. Gall, and A.~Bernstein, ``Tracking concept
  drift of software projects using defect prediction quality,'' in \emph{Mining
  Software Repositories, 2009. MSR'09. 6th IEEE International Working
  Conference on}.\hskip 1em plus 0.5em minus 0.4em\relax IEEE, 2009, pp.
  51--60.

\bibitem[C11]{ethiraj2005capabilities}
S.~K. Ethiraj, P.~Kale, M.~S. Krishnan, and J.~V. Singh, ``Where do
  capabilities come from and how do they matter? a study in the software
  services industry,'' \emph{Strategic management journal}, vol.~26, no.~1, pp.
  25--45, 2005.

\bibitem[C12]{feldgen2012promoting}
M.~Feldgen and O.~Clua, ``Promoting design skills in distributed systems,'' in
  \emph{Frontiers in Education Conference (FIE), 2012}.\hskip 1em plus 0.5em
  minus 0.4em\relax IEEE, 2012, pp. 1--6.

\bibitem[C13]{fogelstrom2008needs}
N.~D. Fogelstr{\"o}m, T.~Gorschek, and M.~Svahnberg, ``Needs oriented framework
  for producing requirements decision material-norm,'' in \emph{Software
  Product Management, 2008. IWSPM'08. Second International Workshop on}.\hskip
  1em plus 0.5em minus 0.4em\relax IEEE, 2008, pp. 9--17.

\bibitem[C14]{hadar2007effective}
E.~Hadar and I.~Hadar, ``Effective preparation for design review: using uml
  arrow checklist leveraged on the gurus' knowledge,'' in \emph{Companion to
  the 22nd ACM SIGPLAN conference on Object-oriented programming systems and
  applications companion}.\hskip 1em plus 0.5em minus 0.4em\relax ACM, 2007,
  pp. 955--962.

\bibitem[C15]{hadar2015cognitive}
I.~Hadar and A.~Zamansky, ``Cognitive factors in inconsistency management,'' in
  \emph{Requirements Engineering Conference (RE), 2015 IEEE 23rd
  International}.\hskip 1em plus 0.5em minus 0.4em\relax IEEE, 2015, pp.
  226--229.

\bibitem[C16]{hunt2015different}
J.~Hunt, ``On a different level of team,'' in \emph{International Conference on
  Agile Software Development}.\hskip 1em plus 0.5em minus 0.4em\relax Springer,
  2015, pp. 254--261.

\bibitem[C17]{kamaludeen2014maintenance}
R.~A. Kamaludeen, Y.-N. Cheah, and S.~Sulaiman, ``Maintenance decision making
  in problem and modification analysis phase using a knowledge-based model,''
  in \emph{Software Engineering Conference (MySEC), 2014 8th Malaysian}.\hskip
  1em plus 0.5em minus 0.4em\relax IEEE, 2014, pp. 25--30.

\bibitem[C18]{knauss2013v}
E.~Knauss and D.~Damian, ``V: issue: lizer: exploring requirements
  clarification in online communication over time,'' in \emph{Proceedings of
  the 2013 International Conference on Software Engineering}.\hskip 1em plus
  0.5em minus 0.4em\relax IEEE Press, 2013, pp. 1327--1330.

\bibitem[C19]{koneri2005design}
P.~G. Koneri, G.-J. de~Vreede, D.~L. Dean, A.~L. Fruhling, and P.~Wolcott,
  ``The design and field evaluation of a repeatable collaborative software code
  inspection process,'' in \emph{International Conference on Collaboration and
  Technology}.\hskip 1em plus 0.5em minus 0.4em\relax Springer, 2005, pp.
  325--340.

\bibitem[C20]{lacerda2009study}
R.~T. Lacerda, L.~Ensslin, and S.~R. Ensslin, ``A study case about a software
  project management success metrics,'' in \emph{Software Engineering Workshop
  (SEW), 2009 33rd Annual IEEE}.\hskip 1em plus 0.5em minus 0.4em\relax IEEE,
  2009, pp. 45--54.

\bibitem[C21]{menzies2009accurate}
T.~Menzies, S.~Williams, O.~Elrawas, D.~Baker, B.~Boehm, J.~Hihn, K.~Lum, and
  R.~Madachy, ``Accurate estimates without local data?'' \emph{Software
  Process: Improvement and Practice}, vol.~14, no.~4, pp. 213--225, 2009.

\bibitem[C22]{okudan20062006}
G.~Okudan, ``2006-1580: An investigation on design effectiveness and efficiency
  of teams equipped with design information support tool (dist),'' \emph{age},
  vol.~11, p.~1, 2006.

\bibitem[C23]{perez2010mobile}
I.~J. P{\'e}rez, F.~J. Cabrerizo, and E.~Herrera-Viedma, ``A mobile decision
  support system for dynamic group decision-making problems,'' \emph{IEEE
  Transactions on Systems, Man, and Cybernetics-Part A: Systems and Humans},
  vol.~40, no.~6, pp. 1244--1256, 2010.

\bibitem[C24]{paech2006open}
B.~Paech and B.~Reuschenbach, ``Open source requirements engineering,'' in
  \emph{Requirements Engineering, 14th IEEE International Conference}.\hskip
  1em plus 0.5em minus 0.4em\relax IEEE, 2006, pp. 257--262.

\bibitem[C25]{rautiainen2003experience}
K.~Rautiainen, L.~Vuornos, and C.~Lassenius, ``An experience in combining
  flexibility and control in a small company's software product development
  process,'' in \emph{Empirical Software Engineering, 2003. ISESE 2003.
  Proceedings. 2003 International Symposium on}.\hskip 1em plus 0.5em minus
  0.4em\relax IEEE, 2003, pp. 28--37.

\bibitem[C26]{roychoudhury2015mining}
S.~Roychoudhury, V.~Kulkarni, and N.~Bellarykar, ``Mining enterprise models for
  knowledgeable decision making,'' in \emph{Proceedings of the Fourth
  International Workshop on Realizing Artificial Intelligence Synergies in
  Software Engineering}.\hskip 1em plus 0.5em minus 0.4em\relax IEEE Press,
  2015, pp. 1--6.

\bibitem[C27]{schubanz2013model}
M.~Schubanz, A.~Pleuss, L.~Pradhan, G.~Botterweck, and A.~K. Thurimella,
  ``Model-driven planning and monitoring of long-term software product line
  evolution,'' in \emph{Proceedings of the Seventh International Workshop on
  Variability Modelling of Software-intensive Systems}.\hskip 1em plus 0.5em
  minus 0.4em\relax ACM, 2013, p.~18.

\bibitem[C28]{siakas2008need}
K.~V. Siakas and E.~Siakas, ``The need for trust relationships to enable
  successful virtual team collaboration in software outsourcing,''
  \emph{International journal of technology, policy and management}, vol.~8,
  no.~1, pp. 59--75, 2008.

\bibitem[C29]{smith1995maintenance}
S.~Smith, K.~Bennett, and C.~Boldyreff, ``Is maintenance ready for evolution?''
  in \emph{Software Maintenance, 1995. Proceedings., International Conference
  on}.\hskip 1em plus 0.5em minus 0.4em\relax IEEE, 1995, pp. 367--372.

\bibitem[C30]{sojer2014understanding}
M.~Sojer, O.~Alexy, S.~Kleinknecht, and J.~Henkel, ``Understanding the drivers
  of unethical programming behavior: The inappropriate reuse of
  internet-accessible code,'' \emph{Journal of Management Information Systems},
  vol.~31, no.~3, pp. 287--325, 2014.

\bibitem[C31]{stray2012investigating}
V.~G. Stray, N.~B. Moe, and A.~Aurum, ``Investigating daily team meetings in
  agile software projects,'' in \emph{Software Engineering and Advanced
  Applications (SEAA), 2012 38th EUROMICRO Conference on}.\hskip 1em plus 0.5em
  minus 0.4em\relax IEEE, 2012, pp. 274--281.

\bibitem[C32]{vahaniitty2005towards}
J.~Vahaniitty and K.~Rautiainen, ``Towards an approach for managing the
  development portfolio in small product-oriented software companies,'' in
  \emph{System Sciences, 2005. HICSS'05. Proceedings of the 38th Annual Hawaii
  International Conference on}.\hskip 1em plus 0.5em minus 0.4em\relax IEEE,
  2005, pp. 314c--314c.

\bibitem[C33]{verbauwhede1995synthesis}
I.~Verbauwhede and J.~M. Rabaey, ``Synthesis for real time systems: Solutions
  and challenges,'' \emph{The Journal of VLSI Signal Processing}, vol.~9,
  no.~1, pp. 67--88, 1995.

\bibitem[C34]{verweij2014user}
P.~Verweij, N.~Marinova, and R.~Lokers, ``User centered design: tools for
  encouraging climate change adaptation,'' 2014.

\end{thebibliography}

\begin{thebibliography}{10}
\providecommand{\url}[1]{#1}
\csname url@samestyle\endcsname
\providecommand{\newblock}{\relax}
\providecommand{\bibinfo}[2]{#2}
\providecommand{\BIBentrySTDinterwordspacing}{\spaceskip=0pt\relax}
\providecommand{\BIBentryALTinterwordstretchfactor}{4}
\providecommand{\BIBentryALTinterwordspacing}{\spaceskip=\fontdimen2\font plus
\BIBentryALTinterwordstretchfactor\fontdimen3\font minus
  \fontdimen4\font\relax}
\providecommand{\BIBforeignlanguage}[2]{{%
\expandafter\ifx\csname l@#1\endcsname\relax
\typeout{** WARNING: IEEEtran.bst: No hyphenation pattern has been}%
\typeout{** loaded for the language `#1'. Using the pattern for}%
\typeout{** the default language instead.}%
\else
\language=\csname l@#1\endcsname
\fi
#2}}
\providecommand{\BIBdecl}{\relax}
\BIBdecl

\bibitem[D1]{arao2014decision}
T.~Arao, Y.~Machida, K.~Toda, R.~Yaegashi, and T.~Takagi, ``Decision-making
  about software release time using analytic hierarchy process,'' in
  \emph{Advanced Applied Informatics (IIAIAAI), 2014 IIAI 3rd International
  Conference on}.\hskip 1em plus 0.5em minus 0.4em\relax IEEE, 2014, pp.
  751--756.

\bibitem[D2]{armour2005project}
P.~G. Armour, ``Project portfolios: organizational management of risk,''
  \emph{Communications of the ACM}, vol.~48, no.~3, pp. 17--20, 2005.

\bibitem[D3]{ballou1996decision}
D.~P. Ballou and G.~K. Tayi, ``A decision aid for the selection and scheduling
  of software maintenance projects,'' \emph{IEEE Transactions on Systems, Man,
  and Cybernetics-Part A: Systems and Humans}, vol.~26, no.~2, pp. 203--212,
  1996.

\bibitem[D4]{bertolino2007performance}
A.~Bertolino, E.~Marchetti, and R.~Mirandola, ``Performance measures for
  supporting project manager decisions,'' \emph{Software Process: Improvement
  and Practice}, vol.~12, no.~2, pp. 141--164, 2007.

\bibitem[D5]{bose1999wwac}
P.~Bose and X.~Zhou, ``Wwac: Winwin abstraction based decision coordination,''
  in \emph{ACM SIGSOFT Software Engineering Notes}, vol.~24, no.~2.\hskip 1em
  plus 0.5em minus 0.4em\relax ACM, 1999, pp. 127--136.

\bibitem[D6]{chen2006software}
Y.~Chen, G.~C. Gannod, and J.~S. Collofello, ``A software product line process
  simulator,'' \emph{Software Process: Improvement and Practice}, vol.~11,
  no.~4, pp. 385--409, 2006.

\bibitem[D7]{choetkiertikul2010risk}
M.~Choetkiertikul and T.~Sunetnanta, ``A risk assessment model for offshoring
  using cmmi quantitative approach,'' in \emph{Software Engineering Advances
  (ICSEA), 2010 Fifth International Conference on}.\hskip 1em plus 0.5em minus
  0.4em\relax IEEE, 2010, pp. 331--336.

\bibitem[D8]{chroust2004empty}
G.~Chroust, ``The empty chair: uncertain futures and systemic dichotomies,''
  \emph{Systems Research and Behavioral Science}, vol.~21, no.~3, pp. 227--236,
  2004.

\bibitem[D9]{di2014information}
D.~Di~Caprio, F.~J. Santos-Arteaga, and M.~Tavana, ``Information acquisition
  processes and their continuity: transforming uncertainty into risk,''
  \emph{Information Sciences}, vol. 274, pp. 108--124, 2014.

\bibitem[D10]{elsood2014goal}
M.~A.~A. Elsood, H.~A. Hefny, and E.~S. Nasr, ``A goal-based technique for
  requirements prioritization,'' in \emph{Informatics and Systems (INFOS), 2014
  9th International Conference on}.\hskip 1em plus 0.5em minus 0.4em\relax
  IEEE, 2014, pp. SW--18.

\bibitem[D11]{fei1992f}
Z.~Fei and X.~Liu, ``f-cocomo: fuzzy constructive cost model in software
  engineering,'' in \emph{Fuzzy Systems, 1992., IEEE International Conference
  on}.\hskip 1em plus 0.5em minus 0.4em\relax IEEE, 1992, pp. 331--337.

\bibitem[D12]{forte1997managing}
G.~Forte, ``Managing change for rapid development,'' \emph{IEEE Software},
  vol.~14, no.~2, pp. 120--122, 1997.

\bibitem[D13]{gachet2003developing}
A.~Gachet and P.~Haettenschwiler, ``Developing intelligent decision support
  systems: A bipartite approach,'' in \emph{International Conference on
  Knowledge-Based and Intelligent Information and Engineering Systems}.\hskip
  1em plus 0.5em minus 0.4em\relax Springer, 2003, pp. 87--93.

\bibitem[D14]{golpayegani2015making}
D.~Golpayegani and F.~Shams, ``Making a tradeoff between adaptation and
  integration in adaptive service based systems,'' in \emph{Computer Science
  and Software Engineering (CSSE), 2015 International Symposium on}.\hskip 1em
  plus 0.5em minus 0.4em\relax IEEE, 2015, pp. 1--7.

\bibitem[D15]{jedlitschka2004towards}
A.~Jedlitschka and D.~Pfahl, ``Towards comprehensive experience-based decision
  support,'' in \emph{European Conference on Software Process
  Improvement}.\hskip 1em plus 0.5em minus 0.4em\relax Springer, 2004, pp.
  34--45.

\bibitem[D16]{kouskouras2007discrete}
K.~G. Kouskouras and A.~C. Georgiou, ``A discrete event simulation model in the
  case of managing a software project,'' \emph{European Journal of Operational
  Research}, vol. 181, no.~1, pp. 374--389, 2007.

\bibitem[D17]{laplante2005modeling}
P.~A. Laplante and C.~J. Neill, ``Modeling uncertainty in software engineering
  using rough sets,'' \emph{Innovations in Systems and Software Engineering},
  vol.~1, no.~1, pp. 71--78, 2005.

\bibitem[D18]{racheva2010conceptual}
Z.~Racheva, M.~Daneva, and A.~Herrmann, ``A conceptual model of client-driven
  agile requirements prioritization: Results of a case study,'' in
  \emph{Proceedings of the 2010 acm-ieee international symposium on empirical
  software engineering and measurement}.\hskip 1em plus 0.5em minus 0.4em\relax
  ACM, 2010, p.~39.

\bibitem[D19]{redouane2007decision}
A.~Redouane, ``A decision making model for software design,'' in \emph{Systems,
  Man and Cybernetics, 2007. ISIC. IEEE International Conference on}.\hskip 1em
  plus 0.5em minus 0.4em\relax IEEE, 2007, pp. 2687--2692.

\bibitem[D20]{redouane2006decision}
------, ``A decision making model for web applications design,'' in
  \emph{Cognitive Informatics, 2006. ICCI 2006. 5th IEEE International
  Conference on}, vol.~1.\hskip 1em plus 0.5em minus 0.4em\relax IEEE, 2006,
  pp. 646--651.

\bibitem[D21]{ribeiro2016decision}
L.~F. Ribeiro, M.~A. d.~F. Farias, M.~Mendon{\c{c}}a, and R.~O. Sp{\'\i}nola,
  ``Decision criteria for the payment of technical debt in software projects: A
  systematic mapping study,'' \emph{ICEIS 2016}, p. 572, 2016.

\bibitem[D22]{singh2013reliability}
J.~Singh and L.~Maurya, ``Reliability assessment and prediction of open source
  software systems,'' in \emph{Image Information Processing (ICIIP), 2013 IEEE
  Second International Conference on}.\hskip 1em plus 0.5em minus 0.4em\relax
  IEEE, 2013, pp. 6--11.

\bibitem[D23]{singpurwalla1991determining}
N.~D. Singpurwalla, ``Determining an optimal time interval for testing and
  debugging software,'' \emph{IEEE Transactions on Software Engineering},
  vol.~17, no.~4, pp. 313--319, 1991.

\bibitem[D24]{steven2006adopting}
J.~Steven, ``Adopting an enterprise software security framework,'' \emph{IEEE
  Security \& Privacy}, vol.~4, no.~2, pp. 84--87, 2006.

\bibitem[D25]{sullivan2010opportunity}
K.~Sullivan, ``Opportunity-centered software development,'' in
  \emph{Proceedings of the FSE/SDP workshop on Future of software engineering
  research}.\hskip 1em plus 0.5em minus 0.4em\relax ACM, 2010, pp. 369--374.

\bibitem[D26]{vantakavikran2007constructing}
P.~Vantakavikran and N.~Prompoon, ``Constructing a process model for decision
  analysis and resolution on cots selection issue of capability maturity model
  integration,'' in \emph{Computer and Information Science, 2007. ICIS 2007.
  6th IEEE/ACIS International Conference on}.\hskip 1em plus 0.5em minus
  0.4em\relax IEEE, 2007, pp. 182--187.

\bibitem[D27]{wang2006formal}
Y.~Wang and Y.~Yuan, ``The formal economic model of software engineering,'' in
  \emph{Electrical and Computer Engineering, 2006. CCECE'06. Canadian
  Conference on}.\hskip 1em plus 0.5em minus 0.4em\relax IEEE, 2006, pp.
  2385--2388.

\bibitem[D28]{wanyama2005towards}
T.~Wanyama and B.~H. Far, ``Towards providing decision support for cots
  selection,'' in \emph{Electrical and Computer Engineering, 2005. Canadian
  Conference on}.\hskip 1em plus 0.5em minus 0.4em\relax IEEE, 2005, pp.
  908--911.

\bibitem[D29]{ziemer2006decision}
S.~Ziemer, P.~R.~F. Sampaio, and T.~St{\aa}lhane, ``A decision modelling
  approach for analysing requirements configuration trade-offs in
  timeconstrained web application development.'' in \emph{SEKE}, 2006, pp.
  144--149.

\end{thebibliography}

\begin{thebibliography}{10}
\providecommand{\url}[1]{#1}
\csname url@samestyle\endcsname
\providecommand{\newblock}{\relax}
\providecommand{\bibinfo}[2]{#2}
\providecommand{\BIBentrySTDinterwordspacing}{\spaceskip=0pt\relax}
\providecommand{\BIBentryALTinterwordstretchfactor}{4}
\providecommand{\BIBentryALTinterwordspacing}{\spaceskip=\fontdimen2\font plus
\BIBentryALTinterwordstretchfactor\fontdimen3\font minus
  \fontdimen4\font\relax}
\providecommand{\BIBforeignlanguage}[2]{{%
\expandafter\ifx\csname l@#1\endcsname\relax
\typeout{** WARNING: IEEEtran.bst: No hyphenation pattern has been}%
\typeout{** loaded for the language `#1'. Using the pattern for}%
\typeout{** the default language instead.}%
\else
\language=\csname l@#1\endcsname
\fi
#2}}
\providecommand{\BIBdecl}{\relax}
\BIBdecl

\bibitem[E1]{chen1991design}
M.~Chen and Y.~I. Liou, ``The design of an integrated group support
  environment,'' in \emph{System Sciences, 1991. Proceedings of the
  Twenty-Fourth Annual Hawaii International Conference on}, vol.~4.\hskip 1em
  plus 0.5em minus 0.4em\relax IEEE, 1991, pp. 333--342.

\bibitem[E2]{cito2016developer}
J.~Cito, ``Developer targeted analytics: supporting software development
  decisions with runtime information,'' in \emph{Proceedings of the 31st
  IEEE/ACM International Conference on Automated Software Engineering}.\hskip
  1em plus 0.5em minus 0.4em\relax ACM, 2016, pp. 892--895.

\bibitem[E3]{dingsoyr2009we}
T.~Dings{\o}yr, F.~O. Bj{\o}rnson, and F.~Shull, ``What do we know about
  knowledge management? practical implications for software engineering,''
  \emph{IEEE software}, vol.~26, no.~3, pp. 100--103, 2009.

\bibitem[E4]{ejnioui2012software}
A.~Ejnioui, C.~E. Otero, and A.~A. Qureshi, ``Software requirement
  prioritization using fuzzy multi-attribute decision making,'' in \emph{Open
  Systems (ICOS), 2012 IEEE Conference on}.\hskip 1em plus 0.5em minus
  0.4em\relax IEEE, 2012, pp. 1--6.

\bibitem[E5]{fiebig1998daisy}
C.~Fiebig and C.~Hayes, ``Daisy: a design methodology for experience-centered
  planning support systems,'' in \emph{Systems, Man, and Cybernetics, 1998.
  1998 IEEE International Conference on}, vol.~1.\hskip 1em plus 0.5em minus
  0.4em\relax IEEE, 1998, pp. 920--925.

\bibitem[E6]{franke2014distribution}
U.~Franke, H.~Holm, and J.~K{\"o}nig, ``The distribution of time to recovery of
  enterprise it services,'' \emph{IEEE Transactions on Reliability}, vol.~63,
  no.~4, pp. 858--867, 2014.

\bibitem[E7]{glazewski2005risk}
S.~R. Glazewski, ``Risk management (is not) for dummies,'' DTIC Document, Tech.
  Rep., 2005.

\bibitem[E8]{hans2013modeling}
R.~T. Hans and E.~Mnkandla, ``Modeling software engineering projects as a
  business: A business intelligence perspective,'' in \emph{AFRICON,
  2013}.\hskip 1em plus 0.5em minus 0.4em\relax IEEE, 2013, pp. 1--5.

\bibitem[E9]{jelassi1987integrated}
M.~T. Jelassi and R.~A. Beauclair, ``An integrated framework for group decision
  support systems design,'' \emph{Information \& Management}, vol.~13, no.~3,
  pp. 143--153, 1987.

\bibitem[E10]{kalumbilo2014linking}
M.~Kalumbilo and A.~Finkelstein, ``Linking strategy, governance, and
  performance in software engineering,'' in \emph{Proceedings of the 7th
  International Workshop on Cooperative and Human Aspects of Software
  Engineering}.\hskip 1em plus 0.5em minus 0.4em\relax ACM, 2014, pp. 107--110.

\bibitem[E11]{karvonen1998computer}
S.~Karvonen, ``Computer supported changes in project management,''
  \emph{International Journal of Production Economics}, vol.~54, no.~2, pp.
  163--171, 1998.

\bibitem[E12]{kirk2004flexible}
D.~Kirk, ``A flexible software process model,'' in \emph{Software Engineering,
  2004. ICSE 2004. Proceedings. 26th International Conference on}.\hskip 1em
  plus 0.5em minus 0.4em\relax IEEE, 2004, pp. 57--59.

\bibitem[E13]{kuehne2005software}
R.~Kuehne, C.~Wille, and R.~Dumke, ``Software agents using simulation for
  decision-making,'' \emph{ACM SIGSOFT Software Engineering Notes}, vol.~30,
  no.~1, p.~5, 2005.

\bibitem[E14]{kuwana1996computer}
E.~Kuwana, E.~Yana, Y.~Sakamoto, Y.~Nakamura, and K.~Horikawa,
  ``Computer-supported meeting environment for collaborative software
  development,'' \emph{Information and Software Technology}, vol.~38, no.~3,
  pp. 221--228, 1996.

\bibitem[E15]{lami2012measuring}
G.~Lami and L.~Buglione, ``Measuring software sustainability from a
  process-centric perspective,'' in \emph{Software Measurement and the 2012
  Seventh International Conference on Software Process and Product Measurement
  (IWSM-MENSURA), 2012 Joint Conference of the 22nd International Workshop
  on}.\hskip 1em plus 0.5em minus 0.4em\relax IEEE, 2012, pp. 53--59.

\bibitem[E16]{larreina2006information}
S.~Larreina, S.~Hernando, and D.~Grisale{\~n}a, ``Information visualization for
  the taking of decisions,'' \emph{WIT Transactions on Information and
  Communication Technologies}, vol.~37, 2006.

\bibitem[E17]{Mallick1999}
S.~Mallick and S.~Krishna, ``Requirements engineering: problem domain knowledge
  capture and the deliberation process support,'' in \emph{Database and Expert
  Systems Applications, 1999. Proceedings. Tenth International Workshop
  on}.\hskip 1em plus 0.5em minus 0.4em\relax IEEE, 1999, pp. 392--397.

\bibitem[E18]{marjaie2010recognition}
S.~A. Marjaie and V.~Kulkarni, ``Recognition of hidden factors in requirements
  prioritization using factor analysis,'' in \emph{Computational Intelligence
  and Software Engineering (CiSE), 2010 International Conference on}.\hskip 1em
  plus 0.5em minus 0.4em\relax IEEE, 2010, pp. 1--5.

\bibitem[E19]{mauerkirchner1997event}
M.~Mauerkirchner, ``Event based simulation of software development project
  planning,'' in \emph{International Conference on Computer Aided Systems
  Theory}.\hskip 1em plus 0.5em minus 0.4em\relax Springer, 1997, pp. 527--540.

\bibitem[E20]{mckinley2007applying}
P.~K. McKinley, B.~H. Cheng, and C.~A. Ofria, ``Applying digital evolution to
  the development of self-adaptive uls systems,'' in \emph{Software
  Technologies for Ultra-Large-Scale Systems, 2007. ULS'07. International
  Workshop on}.\hskip 1em plus 0.5em minus 0.4em\relax IEEE, 2007, pp. 3--3.

\bibitem[E21]{mili2014semantic}
A.~Mili, A.~Jaoua, M.~Frias, and R.~G.~M. Helali, ``Semantic metrics for
  software products,'' \emph{Innovations in Systems and Software Engineering},
  vol.~10, no.~3, pp. 203--217, 2014.

\bibitem[E22]{moreno2010consensus}
J.~Moreno-Rodriguez, F.~Cabrerizo, and E.~Herrera-Viedma, ``A consensus support
  methodology for the initial self-assessment of the efqm excellence model in
  healthcare organisations,'' in \emph{Intelligent Systems Design and
  Applications (ISDA), 2010 10th International Conference on}.\hskip 1em plus
  0.5em minus 0.4em\relax IEEE, 2010, pp. 483--488.

\bibitem[E23]{mustafa2014experimental}
B.~A. Mustafa and A.~Zainuddin, ``An experimental design to compare software
  requirements prioritization techniques,'' in \emph{Computational Science and
  Technology (ICCST), 2014 International Conference on}.\hskip 1em plus 0.5em
  minus 0.4em\relax IEEE, 2014, pp. 1--5.

\bibitem[E24]{nohrer2013c2o}
A.~N{\"o}hrer and A.~Egyed, ``C2o configurator: a tool for guided
  decision-making,'' \emph{Automated Software Engineering}, vol.~20, no.~2, pp.
  265--296, 2013.

\bibitem[E25]{pavur1999software}
R.~Pavur, M.~Jayakumar, and H.~Clayton, ``Software testing metrics: do they
  have merit?'' \emph{Industrial Management \& Data Systems}, vol.~99, no.~1,
  pp. 5--10, 1999.

\bibitem[E26]{pidd2012mixing}
M.~Pidd, ``Mixing other methods with simulation is no big deal,'' in
  \emph{Proceedings of the Winter Simulation Conference}.\hskip 1em plus 0.5em
  minus 0.4em\relax Winter Simulation Conference, 2012, p.~67.

\bibitem[E27]{raffo2003supporting}
D.~M. Raffo and S.-o. Setamanit, ``Supporting software process decisions using
  bi-directional simulation,'' \emph{International Journal of Software
  Engineering and Knowledge Engineering}, vol.~13, no.~05, pp. 513--530, 2003.

\bibitem[E28]{rainsberger2007just}
J.~Rainsberger, ``Just try it,'' \emph{IEEE Software}, vol.~24, no.~6, 2007.

\bibitem[E29]{ramler2004decision}
R.~Ramler, ``Decision support for test management in iterative and evolutionary
  development,'' in \emph{Automated Software Engineering, 2004. Proceedings.
  19th International Conference on}.\hskip 1em plus 0.5em minus 0.4em\relax
  IEEE, 2004, pp. 406--409.

\bibitem[E30]{smith2002project}
J.~Smith, ``Is project management software right for you?'' \emph{Plant
  Engineering-Chicago then Highlands Ranch}, vol.~56, no.~6, pp. 36--38, 2002.

\bibitem[E31]{souri2015new}
A.~Souri and M.~Norouzi, ``A new probable decision making approach for
  verification of probabilistic real-time systems,'' in \emph{Software
  Engineering and Service Science (ICSESS), 2015 6th IEEE International
  Conference on}.\hskip 1em plus 0.5em minus 0.4em\relax IEEE, 2015, pp.
  44--47.

\bibitem[E32]{sukkerd2016multiscale}
R.~Sukkerd, J.~C{\'a}mara, D.~Garlan, and R.~Simmons, ``Multiscale time
  abstractions for long-range planning under uncertainty,'' in \emph{Software
  Engineering for Smart Cyber-Physical Systems (SEsCPS), 2016 IEEE/ACM 2nd
  International Workshop on}.\hskip 1em plus 0.5em minus 0.4em\relax IEEE,
  2016, pp. 15--21.

\bibitem[E33]{wongthongtham2006ontology}
P.~Wongthongtham, E.~Chang, T.~S. Dillon, and I.~Sommerville, ``Ontology-based
  multi-site software development methodology and tools,'' \emph{Journal of
  Systems Architecture}, vol.~52, no.~11, pp. 640--653, 2006.

\end{thebibliography}

\begin{thebibliography}{10}
\providecommand{\url}[1]{#1}
\csname url@samestyle\endcsname
\providecommand{\newblock}{\relax}
\providecommand{\bibinfo}[2]{#2}
\providecommand{\BIBentrySTDinterwordspacing}{\spaceskip=0pt\relax}
\providecommand{\BIBentryALTinterwordstretchfactor}{4}
\providecommand{\BIBentryALTinterwordspacing}{\spaceskip=\fontdimen2\font plus
\BIBentryALTinterwordstretchfactor\fontdimen3\font minus
  \fontdimen4\font\relax}
\providecommand{\BIBforeignlanguage}[2]{{%
\expandafter\ifx\csname l@#1\endcsname\relax
\typeout{** WARNING: IEEEtran.bst: No hyphenation pattern has been}%
\typeout{** loaded for the language `#1'. Using the pattern for}%
\typeout{** the default language instead.}%
\else
\language=\csname l@#1\endcsname
\fi
#2}}
\providecommand{\BIBdecl}{\relax}
\BIBdecl

\bibitem{wced1987ocf}
{World Commission on Environment and Development}, ``Our common future,''
  United Nations, Report, 1987.

\bibitem{penzenstadler2014safety}
B.~Penzenstadler, A.~Raturi, D.~Richardson, and B.~Tomlinson, ``Safety,
  security, now sustainability: The nonfunctional requirement for the 21st
  century,'' \emph{IEEE software}, vol.~31, no.~3, pp. 40--47, 2014.

\bibitem{neumann2012foresight}
P.~G. Neumann, ``The foresight saga, redux,'' \emph{Communications of the ACM},
  vol.~55, no.~10, pp. 26--29, 2012.

\bibitem{becker2016requirements}
C.~Becker, S.~Betz, R.~Chitchyan, L.~Duboc, S.~M. Easterbrook,
  B.~Penzenstadler, N.~Seyff, and C.~C. Venters, ``Requirements: The key to
  sustainability,'' \emph{IEEE Software}, vol.~33, no.~1, pp. 56--65, 2016.

\bibitem{frederick2002time}
S.~Frederick, G.~Loewenstein, and T.~O'donoghue, ``Time discounting and time
  preference: A critical review,'' \emph{Journal of Economic Literature},
  vol.~40, no.~2, pp. 351--401, 2002.

\bibitem{loewenstein2003time}
G.~Loewenstein, D.~Read, and R.~F. Baumeister, \emph{Time and decision:
  Economic and psychological perspectives of intertemporal choice}.\hskip 1em
  plus 0.5em minus 0.4em\relax Russell Sage Foundation, 2003.

\bibitem{biffl2006value}
S.~Biffl, A.~Aurum, B.~Boehm, H.~Erdogmus, and P.~Gr{\"u}nbacher,
  \emph{Value-based software engineering}.\hskip 1em plus 0.5em minus
  0.4em\relax Springer Science \& Business Media, 2006.

\bibitem{keeney1993decisions}
R.~L. Keeney and H.~Raiffa, \emph{Decisions with multiple objectives:
  preferences and value trade-offs}.\hskip 1em plus 0.5em minus 0.4em\relax
  Cambridge university press, 1993.

\bibitem{von1944theory}
J.~Von~Neumann and O.~Morgenstern, \emph{Theory of games and economic
  behavior}.\hskip 1em plus 0.5em minus 0.4em\relax Princeton University Press,
  1944.

\bibitem{tversky1986rational}
A.~Tversky and D.~Kahneman, ``Rational choice and the framing of decisions,''
  \emph{Journal of business}, pp. S251--S278, 1986.

\bibitem{klein1999sources}
G.~Klein, ``Sources of power: How people make decisions','' 1998.

\bibitem{lenberg2014towards}
P.~Lenberg, R.~Feldt, and L.-G. Wallgren, ``Towards a behavioral software
  engineering,'' in \emph{Proceedings of the 7th international workshop on
  cooperative and human aspects of software engineering}.\hskip 1em plus 0.5em
  minus 0.4em\relax ACM, 2014, pp. 48--55.

\bibitem{lenberg2015behavioral}
P.~Lenberg, R.~Feldt, and L.~G. Wallgren, ``Behavioral software engineering: A
  definition and systematic literature review,'' \emph{Journal of Systems and
  Software}, vol. 107, pp. 15--37, 2015.

\bibitem{hofman2011behavioral}
R.~Hofman, ``Behavioral economics in software quality engineering,''
  \emph{Empirical Software Engineering}, vol.~16, no.~2, pp. 278--293, 2011.

\bibitem{betz2015sustainability}
S.~Betz, C.~Becker, R.~Chitchyan, L.~Duboc, S.~Easterbrook, B.~Penzenstadler,
  N.~Seyff, and C.~Venters, ``Sustainability debt: A metaphor to support
  sustainability design decisions,'' in \emph{Proceedings of the Fourth
  International Workshop on Requirements Engineering for Sustainable Systems},
  Ottawa, Canada, 2015.

\bibitem{kitchenham2004procedures}
B.~Kitchenham, ``Procedures for performing systematic reviews,'' Keele
  University, Joint Technical Report, 2004.

\bibitem{ieee2013swebok}
\BIBentryALTinterwordspacing
{IEEE}, \emph{Software engineering body of knowledge (SWEBOK)}.\hskip 1em plus
  0.5em minus 0.4em\relax IEEE, 2013. [Online]. Available:
  \url{http://www.swebok.org}
\BIBentrySTDinterwordspacing

\bibitem{cartwright2014behavioral}
E.~Cartwright, \emph{Behavioral economics}.\hskip 1em plus 0.5em minus
  0.4em\relax Routledge, 2014, vol.~22.

\bibitem{cooper1997blackwell}
C.~L. Cooper and C.~Argyris, \emph{The Blackwell encyclopedia of
  management}.\hskip 1em plus 0.5em minus 0.4em\relax Blackwell, 1997.

\bibitem{ernst2015measure}
N.~A. Ernst, S.~Bellomo, I.~Ozkaya, R.~L. Nord, and I.~Gorton, ``Measure it?
  manage it? ignore it? software practitioners and technical debt,'' in
  \emph{Proceedings of the 2015 10th Joint Meeting on Foundations of Software
  Engineering}.\hskip 1em plus 0.5em minus 0.4em\relax ACM, 2015, pp. 50--60.

\bibitem{zannier2007model}
\BIBentryALTinterwordspacing
C.~Zannier, M.~Chiasson, and F.~Maurer, ``A model of design decision making
  based on empirical results of interviews with software designers,''
  \emph{Information and Software Technology}, vol.~49, no.~6, pp. 637--653,
  Jun. 2007. [Online]. Available:
  \url{//www.sciencedirect.com/science/article/pii/S0950584907000122}
\BIBentrySTDinterwordspacing

\end{thebibliography}
\end{document}